\documentclass[12pt]{article}
\usepackage{amssymb,amsmath}
\include{epsf}

\usepackage{graphicx}
\usepackage[arrow,matrix,curve]{xy}

\usepackage{amsmath}
\usepackage{amssymb}
\usepackage{amscd}
\usepackage{amsfonts}
\usepackage{amsthm}

\newcommand{\e}{{\rm e}}
\newcommand{\ii}{{\rm i}}
\newcommand{\dd}{{\rm d}}
\newcommand{\eqn}[1]{(\ref{#1})}

\def\appendix#1{\addtocounter{section}{1}\setcounter{equation}{0}
\renewcommand{\thesection}{\Alph{section}}
\section*{
\thesection\protect\indent \parbox[t]{11.715cm} {#1}}
\addcontentsline{toc}{section}{Appendix\thesection\ \ \ #1} }
\newcommand{\real}{{\mathbb R}} 

\newcommand{\be}{\begin{equation}}
\newcommand{\ee}{\end{equation}}
\newcommand{\beq}{\begin{equation}}
\newcommand{\eeq}{\end{equation}}
\newcommand{\bea}{\begin{eqnarray}}
\newcommand{\eea}{\end{eqnarray}}
\def\beqa{\begin{eqnarray}}
\def\eeqa{\end{eqnarray}}
\def\nn{\nonumber}
\newcommand{\del}{\partial}

\newcommand{\eq}{\begin{equation}}
\newcommand{\eqa}{\begin{eqnarray}}
\newcommand{\en}{\end{equation}}
\newcommand{\ena}{\end{eqnarray}}

\def\sk{\vskip .4cm}
\def\noi{\noindent}
\def\nn{\nonumber}
\def\epsi{{\varepsilon}}

\newcommand{\al}{\alpha}

\newcommand{\ga}{\gamma}

\newcommand{\de}{\delta}

\newcommand{\RR}{{\mathcal R}}

\newcommand{\R} {R}
\newcommand{\oR}{{\bar{\R}}}

\newcommand{\st}{{\star}}

\newcommand{\FF}{\mathcal F}
\newcommand{\A}{{\mathcal  A}}

\newcommand{\La}{\Lambda }

\newcommand{\D}{\Delta}

\newcommand{\TT}{{\mathcal T}}

\newcommand{\x}{\mathsf{x}}
\newcommand{\y}{\mathsf{y}}

\newcommand{\ff}{{\sf f}}

\newcommand{\rr}{{\R}}

\begin{document}
\begin{titlepage}
\begin{flushright}

\baselineskip=12pt
DISTA-UPO/07\\
DSF--24--2007\\
\hfill{ }\\
\end{flushright}

\begin{center}

\baselineskip=24pt

{\Large\bf Twisting all the way:}\\{\large\bf from Classical
Mechanics to Quantum Fields}

\baselineskip=14pt

\vspace{1cm}

{\bf Paolo Aschieri$^{1}$, Fedele Lizzi$^{2}$ and Patrizia
Vitale$^{2}$}
\\[6mm]
 $^{1}${\it Centro Studi e Ricerche ``Enrico Fermi'' Compendio Viminale, 00184 Roma, Italy}\\
 and  {\it Dipartimento di Scienze e Tecnologie
 Avanzate, Universit\`{a} del
 Piemonte Orientale,}\\ and {\it INFN, Sezione di Torino}\\
{\it Via Bellini 25/G 15100 Alessandria,, Italy}\\$^{2}${\it
Dipartimento di Scienze Fisiche, Universit\`{a}
di Napoli {\sl Federico II}}\\ and {\it INFN, Sezione di Napoli}\\
{\it Monte S.~Angelo, Via Cintia, 80126 Napoli, Italy}\\{\small\tt
aschieri@to.infn.it, fedele.lizzi@na.infn.it,
patrizia.vitale@na.infn.it}
\\[10mm]

\end{center}

\vskip 2 cm

\begin{abstract}
We discuss the effects that a noncommutative geometry induced by a
Drinfeld twist has on physical theories.  
We systematically deform all products and symmetries of the theory. 
We discuss noncommutative classical mechanics, in particular its deformed
Poisson bracket and hence time evolution and symmetries. The twisting is then 
extended to classical fields, and then to the main interest of this work: 
quantum fields. 
This leads to a geometric formulation of quantization on noncommutative 
spacetime, i.e. we establish a noncommutative correspondence principle from
$\st$-Poisson brackets to $\st$-commutators.
In particular commutation relations among creation and annihilation 
operators are deduced. 

\end{abstract}

\end{titlepage}

\section{Introduction}

One of the most interesting and promising fields of research in
theoretical physics is the issue of spacetime structure in
extremal energy regimes. There are evidences from General
Relativity, string theory and black hole physics which support the
hypothesis of a noncommutative structure.  The simplest and
probably most suggestive argument which points at a failure of the
classical spacetime picture at high energy scales comes from the
attempt of conjugating the principles of Quantum Mechanics with
those of General Relativity (see \cite{DoplicherFredenhagenRoberts},
and for a review \cite{doplicher}). If
one tries to locate an event with a spatial accuracy comparable
with the Planck length, spacetime uncertainty relations
necessarily emerge. In total analogy with Quantum Mechanics,
uncertainty relations are naturally implied by the presence of
noncommuting coordinates,
\be
[\hat x^\mu, \hat x^\nu]= \ii \Theta^{\mu\nu} \label{ncgen}
\ee
where $\Theta^{\mu\nu}$ is in general coordinate dependent and its
specific form qualifies the kind of noncommutativity. Therefore,
below Planck length the usual description of spacetime as a
pseudo-Riemannian manifold locally modeled on Minkowski space is not
adequate anymore, and it has been proposed that it be described by
a \emph{Noncommutative Geometry} \cite{connes,landi,ticos}. This
line of thought has been pursued since the early days of Quantum
Mechanics \cite{heisenberg}, and more recently
in \cite{madore}-\cite{Steinacker:2007dq} (see also the recent
review \cite{szabo1}). 
\sk
In this context two relevant issues are the formulation of General 
Relativity and the quantization of field theories on noncommutative 
spacetime. There are different proposals for this second issue, and 
different canonical commutation relations have been considered in the 
literature \cite{FioreSchupp}-\cite{fiorewess}. We here frame this issue in a geometric context and address
it by further developing the twist techniques used in 
\cite{ADMW, ABDMSW, AschieriCorfu} in order to formulate a noncommutative 
gravity theory. 
We see how noncommutative spacetime induces a noncommutative  
phase space geometry, equipped with a deformed Poisson bracket. 
This leads to canonical quantization of fields on noncommutative space. 
\sk
We work in the deformation quantization context; noncommutativity
is obtained by introducing a $\st$-product on the algebra of
smooth functions on spacetime. The most widely studied form of
noncommutativity is the one for which the quantity
$\Theta^{\mu\nu}$ of~\eqn{ncgen} is a constant. This
noncommutativity is obtained trought the Gr\"onewold-Moyal-Weyl
$\star$-product (for a review see \cite{Szaboreview}). 
The product between functions (fields) is given by
\be
\left(f\star h\right)(x) = {\rm exp} \left (\frac{\ii}{2}
\theta^{\mu\nu}\frac{\del}{\del x^\mu} \frac{\del}{\del y^\nu}\right) 
f(x)h(y)\big|_{x=y}
\label{moyal}
\ee
with the $\theta^{\mu\nu}$-matrix constant and antisymmetric. In
particular the coordinates satisfy the relations
\be
x^\mu \star x^\nu -  x^\nu \star x^\mu= \ii\theta^{\mu\nu}
\label{cc}~.
\ee

There are two approaches to study the symmetries (e.g. Poincar\'e symmetry) 
of this noncommutative space. 
One can consider $\theta^{\mu\nu}$ as a 
covariant tensor (see for
example \cite{vega,alvarezmeyervazquez}), 
then the Moyal product is fully covariant under Poincar\'e 
(indeed linear affine) transformations. Poincar\'e symmetry is 
spontaneously broken by the nonzero values $\theta^{\mu\nu}$.
The other approach is to consider the matrix components
$\theta^{\mu\nu}$ as 
fundamental physical constants, like $\hbar$ or $c$. Since the commutator
$x^\mu \star x^\nu -  x^\nu \star x^\mu$ in  ~\eqn{cc} is not 
Lorentz invariant, the usual notion of Poincar\'e symmetry is lost.
However there is still a symmetry, due to a twisted Poincar\'e group \cite{wess,CKNT, CPT, Oeckl},
a quantum Poincar\'e Lie algebra and Lie group invariance that implies that fields on 
noncommutative space are organized according to the same particle 
representations as in commutative space. 

We adopt this second approach and we consider the quantum Lie algebras of
vectorfields on noncommutative spacetime, and of vectorfields on
the noncommutative phase spaces associated to this spacetime,
the quantum Lie algebra of symplectic transformations, and 
that of the constants of motion of a given Hamiltonian system.
These noncommutative spaces and symmetries are obtained by deforming 
the usual ones via a Drinfeld twist \cite{drinfeld}.
For example the Drinfeld twist that implements the Moyal-Weyl 
noncommutativity \eqn{moyal} is 
$\mathcal{F}=\e^{-\frac{\ii}{2}\theta^{\mu\nu}\del_\mu\otimes\del_\nu}$.
\sk
%
%
%

In Section 2 we introduce the twist
$\mathcal{F}=\e^{-\frac{\ii}{2}\theta^{\mu\nu}
\del_\mu\otimes\del_\nu}$
and, starting from the principle that
every product, and in general every bilinear map, 
is consistently deformed by composing it with the appropriate realization of 
the twist $\mathcal{F}$, we briefly review the construction of noncommutative 
 space-time differential geometry as in \cite{ADMW,ABDMSW,AschieriCorfu}.
Vectorfields 
have a  natural $\st$-action on the noncommutative algebras of functions 
and tensorfields, giving rise to the concept of deformed derivations.
These $\st$-derivations form a quantum Lie algebra. In this way we
consider the $\st$-Lie algebra of infinitesimal diffeomorphisms.

In Section 3 we study Hamiltonian mechanics on noncommutative space. 
The differential geometry of phase space is naturally induced from that 
of space-time  (see Sec.2). The twist gives a noncommutative algebra of 
observables and here too we have the $\st$-Lie algebra of vectorfields. 
A $\st$-Poisson bracket is introduced so that the $\st$-algebra of
observables becomes a $\st$-Lie algebra. It can be seen as the 
$\st$-Lie subalgebra of Hamiltonian vectorfields (canonical transformations).
Time evolution is discussed. In particular, constants of motion of translation 
invariant Hamiltonians generate symmetry transformations, they
close a $\st$-Lie symmetry algebra.  Moreover in Section 3.2 we formulate the 
general consistency condition between twists and $\st$-Poisson brackets 
(later applied in Section 4).
In Subsection 3.3.1 we study the deformed symmetries of the harmonic 
oscillator, as well as a deformed harmonic oscillator that conserves usual 
angular momentum.

In Section 4 we generalize the twist setting to the case of an 
infinite number of degrees of freedom. We lift the action of the twist
from functions on spacetime to functionals, and study their $\st$-product 
(in particular a well defined definition of  
$\Phi(x)\st\Phi(y)$ and $a(k)\st a(k')$ is given).
We study the algebra of observables (functionals on phase space), 
and field theory in the Hamiltonian formalism. 
Our inspiring principle is that, having a precise
notion of $\st$-derivation and of $\st$-Lie algebra, 
as in the point mechanics case, we are able
to define a $\st$-Poisson bracket for functionals which is unambiguous 
and which gives the $\st$-algebra of observables a $\st$-Lie algebra structure.
In particular we obtain the $\st$-Poisson bracket between 
canonically conjugated fields.

In Section 5 we similarly deform the algebra of quantum observables
by lifting the action of the twist to operator valued functionals
on space-time. We thus obtain a deformed $\hbar$-noncommutativity 
for operator valued functionals, which is in general nontrivial.
Starting from the usual canonical quantization map for field 
theories on commutative spacetime, $\Phi\stackrel{\hbar}
{\rightarrow}\hat\Phi$, 
we uniquely obtain a quantization scheme for 
field theories on  noncommutative spacetime, and show that it 
satisfies a correspondence principle between $\st$-Poisson brackets and $\st$-commutators.
Finally in order to compare our results with the existing literature 
\cite{FioreSchupp}-\cite{fiorewess} we specialize them to the 
algebra of creation and annihilation operators of noncommutative quantum field 
theory. 
\sk
Throughout this paper we consider just
space noncommutativity, this restriction is in order to have a 
simple presentation of the Hamiltonian formalism.

\section{Twist\label{sectwist}}
\setcounter{equation}{0}
In this section we introduce the concept of twist, and develop
some of the noncommutative geometry associated to it. For the sake of
simplicity we start and concentrate on the twist which gives rise
to the Moyal $\st$-product ~\eqn{moyal}, so that we deform the
algebra of smooth functions $C^\infty(\real^d)$ on space (or
spacetime) $\real^d$. However the results presented hold for a
general smooth manifold and a general twist $\FF$ \cite{ADMW}.
Only formulae with explicit tensor indices
$\mu,\nu...$ in the frame $\del_\mu$ hold exclusively for
the Moyal twist. Comments on the case of a general twist are
inserted in the appropriate places throughout the paper. \sk

The Moyal $\st$-product (\ref{moyal}) between functions can be
obtained from the usual pointwise product $(fg)(x)=f(x)g(x)$ via
the action of a twist operator $\FF$
\eq\label{starprodf}
f\st g:=\mu\circ \FF^{-1}(f\otimes g)~,
\en
where $\mu$ is the usual pointwise product between functions,
$\mu(f\otimes g)=fg$, and the twist operator and its inverse are
\eq
\label{MWTW}
\FF=\e^{-\frac{\ii}{2}\theta^{\mu\nu}\frac{\partial}{\partial
x^\mu} \otimes\frac{\partial}{\partial x^\nu}}~,~~~
\FF^{-1}=\e^{\frac{\ii}{2}\theta^{\mu\nu}\frac{\partial}{\partial
x^\mu} \otimes\frac{\partial}{\partial x^\nu}}~; 
\en 
here $\frac{\partial}{\partial
x^\mu}$ and $\frac{\partial}{\partial
x^\nu}$ are globally defined vectorfields on 
$\real^d$ (infinitesimal translations).  
Given the Lie algebra $\Xi$ of vectorfields with the usual Lie bracket
\be
[u,v]:= (u^\mu \partial_\mu v^\nu)
\partial_\nu-(v^\nu\partial_\nu u^\mu)\partial_\mu~,
\ee
and its universal enveloping algebra $U\Xi$, the twist $\FF$ is an element of $ U\Xi\otimes
U\Xi.~$ The elements of $U\Xi$ are
sums of products of vectorfields, with the identification $uv-vu=[u,v]$.

We shall frequently write (sum over $\al$ understood)
\eq\label{Fff}
\FF=\ff^\al\otimes\ff_\al~~~,~~~~\FF^{-1}=\bar\ff^\al\otimes\bar\ff_\al~,
\en
so that
\eq\label{fhfg}
f\st g:=\bar\ff^\al(f)\bar\ff_\al(g)~.
\en
Explicitly we have
\eq
\FF^{-1}=\e^{\frac{\ii}{2}\theta^{\mu\nu}\frac{\partial}{
\partial x^\mu} \otimes\frac{\partial}{\partial x^\nu}} =\sum
\frac{1}{n!}\left( \frac{\ii}
2\right)^n\theta^{\mu_1\nu_1}\ldots\theta^{\mu_n\nu_n}
\partial_{\mu_1}\ldots\partial_{\mu_n}\otimes
{}\partial_{\nu_1}\ldots\partial_{\nu_n}=\bar\ff^\al\otimes\bar\ff_\al~,\label{faexp}
\ee
so that $\al$ is a multi-index. We also introduce the universal
$\RR$-matrix \eq \RR:=\FF_{21}\FF^{-1}~\label{defUR} \en where by
definition $\FF_{21}=\ff_\al\otimes \ff^\al$. In the sequel we use
the notation \eq
\RR=\R^\al\otimes\R_\al~~~,~~~~~~\RR^{-1}=\oR^\al\otimes\oR_\al~.
\en In the present case we simply have $\RR=\FF^{-2}$ but for more
general twists this is no more the case. The $\RR$-matrix measures
the noncommutativity of the $\star$-product. Indeed it is easy to
see that \eq\label{Rpermutation} h\st g=\oR^\al(g)\st\oR_\al(h)~.
\en The permutation group in noncommutative space is naturally
represented by $\RR$. Formula (\ref{Rpermutation}) says that the
$\st$-product is $\RR$-commutative in the sense that if we permute
(exchange) two functions using the $\RR$-matrix action then the
result does not change. \sk \noi{\sl Note 1:  The class of
$\st$-products that can be obtained from a twist $\FF$ is quite
rich, (for example we can obtain star products that give  the
commutation relations $x\st y=q y \st x$ with $q\in {\mathbb C}$ in two or more
dimensions). Moreover we can consider twists and $\st$-products on
arbitrary manifolds not just on $\real^d$. For example, given a
set of mutually commuting vectorfields $\{X_a\}$  ($a=1,2,\ldots
n$) on a $d$-dimensional manifold $M$, we can consider the twist
\eq\label{betternotation} \FF=\e^{-\frac{\ii}{2}\theta^{a b}{X_a}
\otimes X_b}~. \en
Another example is $\FF=\e^{\frac{1}{2}H\otimes \ln(1+\lambda E)}$
where the vectorfields $H$ and $E$ satisfy $[H,E]=2E$. In these
cases too the $\st$-product defined via (\ref{starprodf}) is
associative and properly normalized.
 In general an element $\FF$ of $U\Xi\otimes U\Xi$ is a twist if it is invertible,  satisfies a cocycle
condition and is properly normalized \cite{drinfeld} (see \cite{AschieriCorfu,ADMW} for a short introduction; see also the book \cite{Majidbook}). The cocycle and the
normalization conditions imply associativity of the $\st$-product
and the normalization $h\st 1=1\st h=h$.}

\subsection{Vectorfields and Tensorfields}
We now use the twist to deform the spacetime commutative geometry
into a noncommutative one. The guiding principle is the
one used to deform the product of functions into the $\st$-product
of functions. Every time we have a bilinear map \be \mu\,: X\times
Y\rightarrow Z~~~~~~~~~~~~~~~~~~\ee where $X,Y,Z$ are
vectorspaces, and where there is an action of $\FF^{-1}$ on $X$
and $Y$ we can combine this map with the action of the twist. In
this way we obtain a deformed version $\mu_\st$ of the initial
bilinear map $\mu$: \eqa \mu_\st:=\mu\circ
\FF^{-1}~,\label{generalpres}&~~~~~~~~~~~~~~& \ena {\vskip -.8cm}
\eqa
{}~~~~~~~~~~~~~\mu_\st\,:X\times  Y&\rightarrow& Z\nn\\
(\x, \y)\,\, &\mapsto&
\mu_\st(\x,\y)=\mu(\bar\ff^\al(\x),\bar\ff_\al(\y))\nn~. \ena The
$\st$-product on the space of functions is recovered setting
$X=Y={\mathcal A}=\mathrm{Fun}(M)$. We now study the case of vectorfields, 1-forms and
tensorfields. \sk \noi {\it Vectorfields $\Xi_\st$}. We deform the
product $\mu : \mathcal{A}\otimes \Xi\rightarrow \Xi$ between the space
$\mathcal{A}=\mathrm{Fun}(M)$ of functions on spacetime $M$ and
vectorfields. A generic vectorfield is $v=v^\nu\partial_\nu$.
Partial derivatives act on vectorfields via the Lie derivative
action \eq\label{onlyconst}
\partial_\mu(v)=[\partial_\mu,v]=\partial_\mu(v^\nu)\partial_\nu~.
\en
According to (\ref{generalpres}) the product
$\mu : \mathcal{A}\otimes \Xi\rightarrow \Xi$
is deformed into the product
\eq
h\st v=\bar\ff^\al(h) \bar\ff_\al(v)~. \en Since
$\FF^{-1}=\e^{\frac{\ii}{2}\theta^{\mu\nu}\partial_\mu \otimes
\partial_\nu}$, iterated use of (\ref{onlyconst}) (e.g.
$\partial_\rho\partial_\mu (v)=
\partial_\rho(\partial_\mu (v))=[\partial_\rho,[\partial_\mu,v]]\,$),
 gives \eq\label{vectstfunc}
h\st v=\bar\ff^\al(h) \bar\ff_\al(v)
=\bar\ff^\al(h)\bar\ff_\al(v^\nu) \partial_\nu=
(h\st v^\nu)\partial_\nu ~.
\en In particular we have
\be
v^\mu\st\partial_\mu=v^\mu\partial_\mu. \label{starvec}
\ee

From (\ref{vectstfunc}) it is easy to see that  $h\st (g\st
v)=(h\st g)\st v$, i.e. that the $\st$-multiplication between
functions and vectorfields is consistent with the $\st$-product
of functions. We denote the space of vectorfields with this
$\st$-multiplication  by $\Xi_\st$. As vectorspaces
$\Xi=\Xi_\st$, but $\Xi$ is an ${\mathcal A}$-module while $\Xi_\st$ is an
${\mathcal A}_\st$-module.

\sk \noi {\it 1-forms $\Omega_\st$}. Analogously, we deform the
product $\mu : {\cal A}\otimes \Omega\rightarrow \Omega$ between the
space $\mathcal{A}=\mathrm{Fun}(M)$ of functions on spacetime $M$ and
1-forms. A generic 1-form is $\rho=\rho_\nu \dd x^\nu$. As for
vectorfields we have
\be
h\st \rho=\bar\ff^\al(h) \bar\ff_\al(\rho).
\ee
The action of $\bar\ff_\al$ on forms is given by iterating the
Lie derivative action of the vectorfield $\del_\mu$ on forms.
Explicitely, if $\rho= \rho_\nu \dd x^\nu $ we have
\be
\del_\mu (\rho) = \del_\mu(\rho_\nu) \dd x^\nu
\ee
and \be \rho=\rho_\nu \dd x^\nu = \rho_\nu \st \dd x^\nu.
\label{funfor}
\ee
 Forms  can
be multiplied by functions from the left or from the right (they are a $\mathcal{A}$ bimodule). If we deform the
multiplication from the right we obtain the new product
\be
 \rho \st h =\bar\ff^\al(\rho) \bar\ff_\al(h)~,
\ee
and we move $h$ to the left
 with the help of the $\RR$-matrix,
\eq
 \rho \st h = \bar \R^\al(h)\st \bar \R_\al(\rho) 
\ee

\sk \noi {\it Tensorfields {$\TT_\st$}}.  Tensorfields form an
algebra with the tensorproduct $\otimes$ (over the algebra of
functions). We define $\TT_\st$ to be the noncommutative
algebra of tensorfields. As vectorspaces $\TT=\TT_\st$; the
noncommutative and associative tensorproduct is obtained by
applying (\ref{generalpres}): \eq\label{defofthetensprodst}
\tau\otimes_\st\tau':=\bar\ff^\al(\tau)\otimes
\bar\ff_\al(\tau')~. \en Here again the action of the twist on
tensors is via the Lie derivative; on vectors we have seen that
it is obtained by iterating (\ref{onlyconst}), on 1-forms it is
similarly obtained by iterating $\del_\mu (h \st \dd g)=
\partial_\mu (h)\st \dd g + h \st \dd \partial_\mu (g)$. Use of the
Leibniz rule gives the action of the Lie derivative on a
generic tensor.

If we consider the local coordinate expression of two tensorfields,
for example of the type
\bea
\tau &=& \tau^{\mu_1,...\mu_m}\partial_{\mu_1}
\otimes_\st\ldots\otimes_\st\partial_{\mu_m} \nonumber\\
\tau' &=&\tau'^{\nu_1,...
\nu_n}\partial_{\nu_1}\otimes_\st\ldots\otimes_\st\partial_{\nu_n}
\eea
then their $\st$-tensor product is
\eq
\tau\otimes_\st\tau'=
\tau^{\mu_1,...\mu_m}\st\tau'^{\nu_1,...\nu_n}
\partial_{\mu_1}\otimes_\st\ldots\otimes_\st\partial_{\mu_m}
\otimes_\st\partial_{\nu_1}\otimes_\st\ldots\otimes_\st\partial_{\nu_n}~.
\en
Notice that since the action of the twist $\FF$ on the partial derivatives $\del_\mu$ is the trivial one,
we have
\be
\partial_{\mu_1}\otimes_\st\ldots\partial_{\mu_n}=\partial_{\mu_1}\otimes\ldots\partial_{\mu_n}.
\ee
There is a natural action of the permutation group on undeformed
arbitrary tensorfields:
\be
\tau\otimes\tau'\stackrel{\sigma}{\longrightarrow}
\tau'\otimes\tau~.
\ee
In the deformed case it is the $\RR$-matrix that provides a
representation of the permutation group on $\st$-tensorfields:
\be
\tau\otimes_\st\tau'\stackrel{\sigma_{_\RR}}{\longrightarrow}
\oR^\al(\tau')\otimes_\st \oR_\al(\tau)~.
\ee
It is easy to check that, consistently with $\sigma_\RR$ being a
representation of the permutation group, we have
$(\sigma_\RR)^2=id$.

Consider now an antisymmetric 2-vector
\be
\La=\frac{1}{2}\La^{ij}(\del_i\otimes\del_j-\del_j\otimes\del_i)=
\frac{1}{2}\La^{ij}\st(\del_i\otimes_\st\del_j-\del_j\otimes_\st\del_i)~.
\ee
Since the action of the ${\mathcal R}$-matrix on the partial derivatives
$\del_\mu$ is the trivial one, we have that $\La$ is both an
antisymmetric 2-vector and a $\st$-antisymmetric one.

\subsection{$\st$-Lie Algebra of Vectorfields}

The $\st$-Lie derivative on the algebra of functions $\mathcal{A}_\st$ is
obtained following the general prescription (\ref{generalpres}).
We combine the usual Lie derivative on functions ${\mathcal
L}_uh=u(h)$ with the twist $\FF$ \eq\label{stliederact} {\mathcal
L}^\st_u(h):=\bar\ff^\al(u)(\bar\ff_\al(h))~.
\en
By recalling that every vectorfield can be written as
$u=u^\mu\st\partial_\mu=u^\mu\partial_\mu$ we have
\eqa\nn
{\mathcal L}^\st_u(h)&=&\bar\ff^\al(u^\mu\partial_\mu)
(\bar\ff_\al(h))=\bar\ff^\al(u^\mu)\,\partial_\mu(\bar\ff_\al(h))\\[.3em]
&=&
u^\mu\st\partial_\mu(h)~,\label{exlu}
\ena
where in the second equality we have considered the explicit
expression~(\ref{faexp}) of $\bar\ff^\al$ in terms of partial
derivatives, and we have iteratively used the property
$[\partial_\nu,u^\mu\partial_\mu]=\partial_\nu(u^\mu)\,\partial_\mu$.
In the last equality we have used that the partial derivatives
contained in $\bar\ff_\al$ commute with the partial derivative
$\partial_\mu$.

The differential operator ${\mathcal L}^\st_u$ satisfies the
deformed  Leibniz rule
\be
{\mathcal L}^\st_u(h\st g)={\mathcal
L}_u^\st(h)\st g + \oR^\al(h)\st {\mathcal
L}_{\oR_\al(u)}^\st(g)~. \label{defL}
\ee
This deformed Leibniz rule is intuitive: in the second addend
we have exchanged the order of $u$ and $h$, and this is
achieved by the action of the $\RR$-matrix, that, as observed,
provides a representation of the permutation group.

The Leibniz rule is consistent (and actually follows) from the
coproduct rule
\be
u\mapsto \Delta_\st u=u \otimes 1 +\bar R^\al\otimes \bar R_\al(u)
\label{cop}
\ee
(this formula holds also for the twist
(\ref{betternotation}). However in the most generic twist case
the term $\oR^\al$ has to be replaced with $\ff^\beta
(\oR^\al)\, \ff_\beta$ \cite{ADMW}).

In the commutative case the commutator of two vectorfields
is again a
vectorfield, we have the Lie algebra of vectorfields. In this
$\st$-deformed case we have a similar situation.
We first calculate
\eq
{\mathcal L}^\st_u{\mathcal L}^\st_v(h)={\mathcal
L}^\st_u({\mathcal L}^\st_v(h)) =u^\mu\st\partial_\mu(v^\nu)\st
\partial_\nu (h)+ u^\mu\st v^\nu\st\partial_\nu\partial_\mu(h)
\nn\en Then instead of considering the composition ${\mathcal
L}^\st_v{\mathcal L}^\st_u$ we consider ${\mathcal
L}^\st_{\oR^\al(v)}{\mathcal L}^\st_{\oR_\al(u)}$ Indeed the
usual commutator is constructed permuting (transposing) the two
vectorfields, and we have just remarked that the action of the
permutation group in the noncommutative case is obtained using the
$\RR$-matrix. We have
\eq
{\mathcal L}^\st_{{\oR^\al}(v)}{\mathcal L}^\st_{\oR_\al(u)}(h)
={\oR^\al}(v^\nu)\st \oR_\al(\partial_\nu u^\mu)\st\partial_\mu
h +{\oR^\al}(v^\nu)\st\oR_\al(u^\mu)\st\partial_\nu
\partial_\mu h \nonumber~.
\ee
In conclusion
\be\label{stLieastrep} {\mathcal L}^\st_u\,{\mathcal L}^\st_v-
{\mathcal L}^\st_{\bar R^\al (v)}\,{\mathcal L}^\st_{\bar
R_\al(u)}={\mathcal L}^\st_{[u,v]_\st}
\ee
where we
have defined the new vectorfield \eq\label{stbrack} [u,v]_\st
:= (u^\mu\st\partial_\mu v^\nu)
\partial_\nu-(\partial_\nu u^\mu\st v^\nu)\partial_\mu~.
\en
A more telling definition
of the $\st$-bracket is
\eq
[u,v]_\st := [\bar\ff^\al(u) , \bar\ff_\al(v)]~,\label{etwa}
\en
again as in (\ref{generalpres})
the deformed bracket is obtained from
the undeformed one via composition with the twist: 
\be[~,~]_\st=[~,~]
\circ\FF^{-1}.\label{etwa2}
\ee
Therefore, in the presence of twisted noncommutativity, we replace the usual Lie algebra of vectorfields, 
$\Xi$, with $\Xi_\st$,  the algebra  of vectorfields 
equipped with the $\st$-bracket \eqn{etwa} or equivalently \eqn{etwa2}.

It is not difficult to see that the bracket
$
[~~,~~]_\st~:~\Xi_\st \times\Xi_\st \rightarrow \Xi_\st\nonumber
$
is a bilinear map and verifies  the $\st$-antisymmetry and the
$\st$-Jacoby identity
\eq\label{sigmaantysymme}
[u,v]_\st =-[\oR^\al(v), \oR_\al(u)]_\st~ .
\en
\eq\label{stJac}
[u,[v,z]_\st ]_\st =[[u,v]_\st ,z]_\st
+ [\oR^\al(v), [\oR_\al(u),z]_\st ]_\st ~.
\en
For example we have 
\[
[u,v]_\st=[\bar\ff^\beta (u),\bar\ff_\beta (v)]
=-[\bar\ff_\beta(v),\bar\ff^\beta(u)]=
[\bar\ff^\de\ff^\ga\bar\ff_\beta(v),\bar\ff_\de\ff_\ga\bar\ff^\beta(u)]=
-[\oR^\al(v),\oR_\al(u)]_\st~.
\]
where in the third passage we inserted $1\otimes 1$ in the form $\FF^{-1}\FF$. 
\sk
We have constructed the deformed Lie algebra of vectorfields
$\Xi_\st$. As vectorspaces $\Xi=\Xi_\st$, but $\Xi_\st$ is a $\st$-Lie algebra.
We stress that a $\st$-Lie algebra is not a generic name for a
deformation of a Lie algebra. Rather it is a quantum Lie algebra
of a quantum (symmetry) group \cite{Woronowicz},
(see \cite{Modave} for a short introduction and further references).
In this respect the deformed Leibniz rule (\ref{defL}), that states that
only vectorfields (or the identity) can act on the second argument
$g$
in $h\st g$
(no higher order differential operators are allowed on $g$) is of fundamental
importance (for example it is  a key ingredient for the definition of a
covariant derivative along a generic vectorfield).

\sk

Usually in the literature concerning twisted symmetries the Hopf algebra 
$U\Xi^\FF$ is considered. This has the same algebra structure as $U\Xi$ so that
the Lie bracket is the undeformed one. Also the action of $U\Xi^\FF$
on functions and tensors is the undeformed one (so that no $\st$-Lie 
derivative ${\mathcal L^\st}$ is introduced). It is the coproduct $\Delta^\FF$ 
of $U\Xi^\FF$ that is  deformed: for all $\xi\in U\Xi$, 
$$\Delta^\FF (\xi)=\FF\Delta(\xi)\FF^{-1}~.$$ The $\st$-Lie algebra
 $\Xi_\st$ we have constructed gives rise to the universal enveloping 
algebra $U\Xi_\st$ of sums of products of vectorfields, 
with the identification $u\st v-\oR^\al(v)\st\oR_\al(u)=[u,v]_\st$ 
and coproduct (\ref{cop}) \cite{ADMW,AschieriCorfu}.  
The Hopf (or symmetry) algebras 
$U\Xi^\FF$ and $U\Xi_\st$ are isomorphic. Therefore to some extent it is a 
matter of taste wich algebra one should use.
We prefer  $U\Xi_\st$ because $U\Xi_\st$ naturally arises from the general 
prescription (\ref{generalpres}): the product $u\st v$ in $U\Xi_\st$ is just
$u\st v=\bar\ff^\al(u)\bar\ff_\al(v)$, and
because it is in $U\Xi_\st$ (not in $U\Xi^\FF$) 
that vectorfields have the geometric meaning of {\sl infinitesimal} 
generators, for example the coproduct $\D_\st(t)$ is a minimal deformation 
of the usual coproduct $\D(t)=t\otimes 1+1\otimes t$. Also, from (\ref{exlu}),
we have the  ${\mathcal A}_\st$-linearity  property
${\mathcal{L}}^\st_{f\st u}h=f\st{\mathcal{L}}^\st_{u} h$.

\section{Classical Mechanics}
\setcounter{equation}{0}

In this section we apply the programme we outlined to classical
mechanics, thus building a \emph{$\st$-classical
mechanics}. A main motivation is the construction of a deformed
Poisson bracket, and the study of its geometry. The Poisson
bracket will be generalized to field theory in the next section.

In subsection 3.1 we briefly review the geometry of usual phase space, 
then we lift the action of the twist $\FF$ from spacetime to phase space.
The structures introduced in Section 2  immediately give the  
differential geometry on noncommutative phase space.  The deformation of the 
standard Poisson bracket on $\real^{2n}$ and the $\st$-Lie algebra of 
Hamiltonian vectorfields are then studied.
The general case of an arbitrary Poisson bracket deformed by an 
arbitrary twist $\FF$ is considered in subsection 3.2, there we see that
a compatibility requirement between the twist $\FF$ and the Poisson bracket 
emerges.

In subsection 3.3 we study Hamiltonian dynamics.  
The constants of motion of translation invariant 
Hamiltonians generate symmetry transformations, and
close a $\st$-Lie subalgebra under the $\st$-Poisson bracket.  
We also study the harmonic oscillator as an example of
noncommutative Hamiltonian dynamics that is not translation invariant.
%
%

\subsection{$\st$-Poisson Bracket}
In the Hamiltonian approach the dynamics of a classical
finite-dimensional mechanical system is defined through a Poisson
(usually symplectic) structure on phase space and the choice of a
Hamiltonian function. The Poisson structure is a bilinear map
\be \{\ ,\
\}:\mathcal A \times\mathcal A \longrightarrow \mathcal A
\label{poisson}
\ee
where $\mathcal A$ is the algebra of smooth functions on phase space.
It satisfies
\beqa
&& \{f,g\}=-\,\{g,f\} ~~~~~~~~~~~~~~~~~~~~~~~~~~~~~~~~~~~~~~~{\mbox{\sl antisymmetry}}\\
&&\{f,\{g, h\}\}  + \{h,\{f,g\}\} + \{g, \{h,f\}\}= 0 ~~~~\,~~{\mbox{\sl Jacobi identity}}\\
&& \{f,g h\}= \{f, g\} h + g\{f, h\}~~~~~~~~~~~~~~~~~~~~~~~~~ {\mbox{\sl Leibniz rule}}
\eeqa
The first two properties show that the Poisson bracket $\{~,~\}$ is a Lie
bracket.
The last property shows that the map $\{f,~\} :
\mathcal A\rightarrow \mathcal A$ is a derivation of the algebra $\mathcal A$,
it therefore defines a vectorfield
\be
X_f := \{f, ~\}~,
\ee
so that $ \{f, g\}=X_f(g)=\langle X_f,\dd g \rangle $.  $X_f$ is
the Hamiltonian vectorfield associated to the
``Hamiltonian'' $f$. We will also use the notation $\{f, ~\}=
\mathcal{L}_{X_f}$ where $\mathcal{L}_{X_f}$ is the Lie
derivative. The antisymmetry property shows that the vector field
$X_f$ actually depends on $f$ only through its differential $\dd
f$, and we thus arrive at the Poisson bivector field $\La$ that
maps 1-forms into vectorfields according to
\be
\langle\Lambda, \dd f\rangle = X_f~.
\ee
We therefore have
\be
\langle\Lambda, \dd f\otimes \dd g\rangle = X_f(g)=\{f,g\}~.\label{onion}
\ee
Notice that we use the pairing $\langle u\otimes v, \dd f\otimes d
g\rangle =\langle v,\dd f\rangle \,\langle u, \dd g\rangle$ ($u$ and
$v$ vectorfields) that is obtained by first contracting the innermost elements. We use this onion-like structure pairing because it naturally generalizes to the
noncommutative case.

\sk
To be definite let us consider the canonical bracket on the
phase space ${\rm T^*}\real^n$ with the usual coordinates
$x^1,\ldots x^n, p_1,\ldots p_n$
\be
\{f,g\}:=\frac{\del f}{\del{x^\ell}}  ~ \frac{\del
g}{\del{p_\ell}}\,-\,\frac{\del f}{\del{p_\ell}} ~ \frac{\del
g}{\del{x^\ell}}  ~,
\ee
sum over repeated indices (which take the values $1,\ldots  n$) is assumed.

Because of the onion like structure of the pairing and since
$\langle \frac{\del }{\del{x^i}} \otimes \frac{\del
}{\del{p_i}}\,,\dd f \rangle= \frac{\del f }{\del{p_i}} \frac{\del
}{\del{x^i}}$, we have that the Poisson bivector field is
\eq
\Lambda = \frac{\del }{\del{p_i}}\,  \wedge \,\frac{\del
}{\del{x^i}} = \frac{\del }{\del{p_i}}  \otimes \frac{\del
}{\del{x^i}} - \frac{\del }{\del{x^i}} \otimes \frac{\del
}{\del{p_i}}  ~,
\ee
while 
\be
X_f= \frac{\del f}{\del{x^i}}  \frac{\del }{\del{p_i}}  -
\frac{\del f}{\del{p_i}} \frac{\del }{\del{x^i}}   ~.
\ee
The symplectic form associated to the nondegenerate Poisson tensor
$\La$ satisfies $\{f,h\}=\langle X_f\otimes X_h , \omega\rangle$
and explicitly reads \eq \omega=dp_i\wedge \dd x^i~. \en

A Hamiltonian $H$ is  a function on phase space. Motion of a
point in phase space describes the time evolution of the dynamical
system. Infinitesimally it is given by the vectorfield $X_H$, and
on the algebra $\A$ of observables (not explicitly dependent on
time), we have Hamilton's equation
\be
\dot{ f} = -\{H, f\} = -{X_H} (f) ~.
\ee



We denote with $\sigma_t$ the integral flow of $-X_H$.
If the system at time $t_0=0$ is described by the point $P_0$ in
phase space, at a later time $t$ is has evolved to the point
$P_t=\sigma_t(P_0)$. Correspondingly the time evolution of any
observable is
\be
\sigma^*_t(f)=f\circ \sigma_t~,
\ee
where $\sigma^*_t$ is the
pull-back of the integral flow. 
In particular the coordinates of the point $P_t$ are
$x^i(t)=x^i(\sigma_t(P_0))$ and $p_i(t)=p_i(\sigma_t(P_0))$.
Hamilton's equation can be equivalently rewritten as an equation for the 
pull-back flow 
$\sigma_t^\st$,
\eq
\frac{\rm d}{{\rm d} t}{\,\sigma_t^\st}= 
- {\sigma_t^*}\circ X_H ~.\label{pbflow}
\en
\sk Now we twist commutative spacetime into noncommutative
spacetime (actually we consider just noncommutative space
coordinates, no time noncommutativity). Correspondingly the
configuration space and the phase space of a mechanical system
will be noncommutative. For example if space is $\real^3$ and we
consider an unconstrained mechanical system of $r$ points then
the configuration space will be $\real^{3r}$. Noncommutativity on
$\real^{3r}$ is induced from noncommutativity on $\real^{3}$.
Recall that $\real^{3r}$ should be considered as $r$ copies of
$\real^{3}$, therefore a transformation on $\real^{3}$ induces a
simultaneous transformation on all the $r$ copies of $\real^{3r}$.
Infinitesimally, if the transformation on $\real^{3}$ (with
coordinates $x^k$, $k=1,2,3$) is given by the vectorfield
$\frac{\del }{\del{x^i}}$, then the corresponding infinitesimal
transformation on $\real^{3r}$ is given by the vectorfield
\be
\frac{\del }{\del{x_1^i}} + \frac{\del }{\del{x_2^i}} \ldots +
\frac{\del }{\del{x_r^i}}
\ee
(with $x^k_1,x^k_2,\ldots x^k_r$ coordinates of $\real^{3r}$). We
therefore have the following lift of the action of the twist
$\FF$ from $C^\infty(\real^3)$ to $C^\infty(\real^{3r})\otimes
C^\infty(\real^{3r})$,
\eq
\mathcal{F}=\ff^\alpha\otimes\ff_\alpha =\e^{-\frac{\ii}2\theta^{ij}({\frac{\del }{\del{x_1^i}}
+ \ldots
\frac{\del }{\del{x_r^i}}})\,\otimes\,
(\frac{\del }{\del{x_1^j}} + \ldots
\frac{\del }{\del{x_r^j}})}~,\label{twistconfspace}
\en
and correspondingly the following $\st$-product on configuration
space,
for all $a,b\in C^\infty(\real^{3r})$,
\eq
a_{}\st_{}b\,(x_1,...x_r)=
{\rm exp}  \left(\frac{\ii}{2}
\theta^{ij}
({\frac{\del }{\del{x_1^i}}
+ \ldots
\frac{\del }{\del{x_r^i}}})\,
(\frac{\del }{\del{y_1^j}} + \ldots
\frac{\del }{\del{y_r^j}})\right)
 \,a(x_1,...x_r)\,b(y_1,... y_r) \big|_{x=y}
\label{lifttoconfspace}
\en
On the subalgebra
$C^\infty(\real^{3})\otimes\ldots \otimes C^\infty(\real^{3})$ ($r$-times)
of $C^\infty(\real^{3r})$ the $\st$-product (\ref{lifttoconfspace})
coincides with the one defined in \cite{fiorewess}.

We further lift the twist $\FF$ to the tangent bundle
T$\,\real^{3r}$ and to the phase space T$^*\real^{3r}$. A point of the
manifold T$\,\real^{3r}\simeq \real^{6r}$ has coordinates
$(x^A,v^A)$, ($A=1,\ldots,3r$) where $v^A$ are the components of
the vector $v=v^A\frac{\del}{\del{x^A}}$ tangent  to the point of
coordinates $x^A$. Under the translation generated by
$({\frac{\del }{\del{x_1^i}} + \ldots \frac{\del }{\del{x_r^i}}})
$ we have that $(x^A,v^A)$ is translated into $(x'^A,v^A)$, where
$x'^A$ are the new coordinates of the translated point, while the
coefficients $v^A$ do not change because we are considering a
constant translation. Therefore the action of $({\frac{\del
}{\del{x_1^i}} + \ldots \frac{\del }{\del{x_r^i}}}) $, and of the
twist $\FF$, on the tangent bundle T$\,\real^{3r}$ is the usual
one on the base space and the trivial one on the fibers. Similarly
for the phase space T$^*\,\real^{3r}$. Let $x^A,p_A$ be phase
space coordinates, the explicit expression of $\FF$ on
$C^\infty(\rm{T}^* \real^{3r})\otimes C^\infty(\rm{T}^*
\real^{3r})$ is again (\ref{twistconfspace}). In particular $f\st
h=fh$ if $f$ or $h$ is only a function of the momenta $p_A$.

\noi{\sl Note 2: This result holds just because of the particular
twist we have considered. In general the lift of a vectorfield
$u=u^A\frac{\del }{\del{x^A}}$ from $\real^{3r}$ to
$\mathrm{T}\,\real^{3r}$ is given by ${u_*}=u^A\frac{\del
}{\del{x^A}}+v^B\frac{\del u^A}{\del{x^B}} \frac{\del
}{\del{v^A}}$ (here $x^A,v^A$ are the coordinates of
$\mathrm{T}\,\real^{3r}$). Notice the linearity of ${u_*}$ in the
fiber coordinates $v^A$, indeed the lift ${u_*}$ can be obtained
from its flow $T\sigma^u_t$, that is linear on the fibers because
it is a tangent flow, precisely the differential of the flow
$\sigma^u_t$ associated to the vector field $u$. Similarly the
lift of $u$ to the phase space  $\mathrm{T}^*\real^{3r}$ (with
coordinates $x^A,p_A$), is given by the vector field
\be u^*=u^B\frac{\del }{\del{x^B}}-p_B\frac{\del u^B
}{\del{x^C}}\frac{\del }{\del{p_C}}~.\ee From these explicit
formulae we see that more general twists, constructed for example
with mutually commuting vectorfields like in
(\ref{betternotation}), act nontrivially on the fibers of the
tangent and contangent bundle.
%
%
%
}

We have seen how noncommutativity of spacetime induces
noncommutativity of phase space. Let us consider a system with $n$ degrees of freedom with phase space 
 $M=\real^{2n}$, and  $\A_\st=C^\infty(M)_\st$ the noncommutative
algebra of functions on $M$ with twist 
\be\label{twisttt} {\mathcal
F}=\e^{-\frac{\ii}{2}\theta^{\ell s}
\frac{\del}{\del{x^\ell}}\otimes
\frac{\del}{\del^{}{x^{s^{}}}}}~~~~~~~~\ell, s =1,...n~.
\ee
It can be easily checked that the  Poisson bracket does not define a  derivation
of the algebra $\A_\st=C^\infty(M)_\st$,
\be
\{f, g\star h\}\ne \{f,g\}\star h + g\star \{f,h\}
\ee
or, in different words,
\eq
\mathcal{L}_{ X_f} (g\star h)\ne (\mathcal{L}_{ X_f} g)\star h+
g\star (\mathcal{L}_{ X_f}  h)~. \label{leibXf}
\ee
On the other hand, according to \eqn{generalpres}, we are 
 led to deform the Poisson structure into a noncommutative
Poisson structure $\{~,~\}_\st$. We define the $\st$-Poisson
bracket
\eq
\{f,g\}_{\star}:=\{\bar\ff^\al(f),\bar\ff_\al(g)\}~.
\ee
A simple calculation, that exploits the fact that the Poisson
structure is invariant under the partial derivatives appearing in
the twist, shows that this twisted Poisson bracket can be
expressed as:
\be\label{explformPoi}
\{f,g\}_{\star}= \frac{\del f}{\del{x^\ell}} \,\st\, \frac{\del
g}{\del{p_\ell}} \,-\, \frac{\del f}{\del{p_\ell}}  \,\st\,
\frac{\del g}{\del{x^\ell}} ~.
\ee
This bracket is linear in
both arguments, it is $\mathcal R$-antisymmetric and it satisfies
the $\st$-Leibniz rule and  $\st$-Jacobi identity:
\beqa
\{f,g\}_{\star}&=&-\{\bar\rr^\alpha(g),\bar
\rr_\alpha(f)\}_{\star}\label{stantis}\\
\{f,g\star h\}_\star &=&\{f,g\}_\star \star h +\bar
\rr^\alpha(g)\star \{\bar \rr_\alpha(f),h\}_\star
\label{leibtwist}\\
\{f,\{g, h\}_\star\}_\star &=&  \{\{f,g\}_\star,h\}_\star +
\{\bar\rr^\alpha(g), \{\bar\rr_\alpha(f),h\}_\star\}_\star
\label{jacobitwist}
\eeqa
We conclude from  (\ref{leibtwist})  that $\{f, ~ \}$ is a
$\star$-derivation. We can write
\eq
 \{f, ~\}_\star = {\mathcal L}_{v}^\st
\en
for some vectorfield $v$. From (\ref{explformPoi})  and the definition of
$\st$-Lie derivative, we deduce that the vectorfield $v$ is the undeformed
Hamiltonian vector field $v=X_f=\{f,~\}$, therefore we obtain
\be
\{f, ~\}_\star={\mathcal L}^\st_{X_f}={\mathcal L}^\st_{\{f,~\}}~. \label{Xf}
\ee
The Leibniz rule (\ref{leibtwist}) can be rewritten as
\eq\label{LeibHam}
{\mathcal L}^\st_{X_f}(g\st h)={\mathcal L}_{X_f}^\st(g)\st h +
\oR^\al(g)\st {\mathcal L}_{X_{\oR_\al(f)}}^\st(h)~
\en
and is consistent (and actually follows) from the coproduct rule
\eq\label{copHam}
X_f\mapsto \Delta_\st X_f=X_f\otimes 1 +\oR^\al\otimes X_{\oR_{\al(f)}}~.
\en

Property (\ref{jacobitwist}), the $\st$-Jacobi identity,  can be rewritten as \eq {\mathcal
L}^\st_{X_f}\,{\mathcal L}^\st_{X_g}- {\mathcal L}^\st_{\oR^\al
(X_g)}\,{\mathcal L}^\st_{\oR_\al(X_f)}={\mathcal
L}^\st_{X_{\{f,g\}_\st}}\;. \en Recalling (\ref{stLieastrep}) we
equivalently have \eq
[X_f,X_g]_\st=X_{\{f,g\}_\st}~.\label{closure} \en Because of this
property and of the Leibniz rule (\ref{LeibHam}) (or better the
coproduct rule (\ref{copHam})) {\sl Hamiltonian vector fields are
a $\star$-Lie subalgebra of the $\star$-Lie algebra of vectorfields}.

\subsection{General Twist and Poisson Bracket} 
These results, obtained in the
case of the $\theta$-constant twist (\ref{twistconfspace}) or
(\ref{twisttt}) on $M=\real^{2n}$, can be generalized to a twist
$\FF$ on an arbitray Poisson manifold $M$ (phase space). We
comment on this general case because it is in this context
that the compatibility relation between twist and Poisson
structure most clearly emerges.
The twist deforms the algebra of functions on $M$ into the $\st$-algebra
$\A_\st=C^\infty_\st(M)$, where $f\st g= \bar\ff^\al(f)\,\bar\ff_\al(g)~.$
According to the general principles we have set in Section \ref{sectwist},
first we define the $\st$-pairing between vectorfields and $1$-forms
\eq
\langle u, \vartheta\rangle_\st := \langle \bar\ff^\al(u) ,
\bar\ff_\al(\vartheta)\rangle~.
\en 
It can be proven that this
pairing has the $\A_\st$-linearity properties 
\be
\langle f\st u ,
\vartheta\st h\rangle_\st = f\st \langle u , \vartheta\rangle_\st \st h
\ee
(where $\vartheta\st h:=\bar\ff^\al(\vartheta) \bar\ff_\al(h)$)
and 
\be
\langle u , f\st\vartheta\rangle_\st = \oR^\al(f)\st
\langle \oR_\al(u) , \vartheta\rangle_\st~.
\ee
We extend the pairing to covariant tensors, $\tau$, and contravariant ones,
$\rho$, via the definition
\eq
\langle \tau, \rho \rangle_\st :=
\langle \bar\ff^\al(\tau) , \bar\ff_\al(\rho)\rangle~.
\en
It can be shown that this definition, and the onion like structure of the
undeformed pairing (cf. after (\ref{onion})), imply
the property
\eq
\langle u\otimes_\st v , \vartheta\otimes_\st \eta\rangle_\st:=
\langle u\,,\,\langle v,\vartheta\rangle_\st\st\eta\rangle_\st~,\label{2vf}
\en
(where $\eta$ is a 1-form). This equation gives an equivalent definition of
the pairing between covariant and contravariant 2-tensors. From (\ref{2vf})
it follows that the $\A_\st$-linearity properties are preserved:
\bea
\langle f\,\st \,u\otimes_\st v , \vartheta\otimes_\st \rho \st
h\rangle_\st &=& f\st \langle u\otimes_\st v ,
\vartheta\otimes_\st \rho\rangle_\st\st h\nonumber\\
\langle u\otimes_\st v , f\,\st\,\vartheta\otimes_\st
\rho\rangle_\st &=& \oR^\al(f)\st \langle \oR_\al(u\otimes_\st v)
, \vartheta\otimes\st \rho\rangle_\st~.
\eea
Finally, following
(\ref{generalpres}), we define the $\st$-Poisson bracket as
\be \label{generaldefP}
\{f,g\}_{\star}:=\langle \La , \dd f\otimes_\st \dd g\rangle_\st~.
\ee
Using the fact that $\La$ is $\st$-antisymmetric the
$\st$-antisymmetry property (\ref{stantis}) can be proven. 
However from the definition
(\ref{generaldefP}) it follows that 
\eq 
\{f, g\st h\}_\st=\{f,
g\}_\st \,\st h + \oR^\al\oR^\beta(g)\,\st\,\langle \oR_\al(\La),
d{\oR_\beta} (f)\otimes_\st dh \rangle_\st ~. 
\en 
This equality
becomes the deformed Leibniz rule (\ref{leibtwist}) if 
\eq
\oR^\al\otimes \oR_\al(\La)= 1\otimes \La \label{compatibility}
\en 
(recall that $1$ and $\oR^\al$ are elements in $U\Xi$). This
is a compatibility relation between the Poisson structure and the
twist.

Led by this observation we require, as compatibility condition,
that the action of the twist $\FF$ on the Poisson tensor $\La$ be
the trivial one,
\eq
\bar\ff^\al\otimes \bar\ff_\al(\La) = 1\otimes \La~,
\label{compLATW1} 
\en
\eq
\bar\ff^\al(\La)\otimes \bar\ff_\al = \La\otimes
1~.\label{compLATW2} 
\en 
Any two of the last three equations imply
the third one. If we consider a twist of the form
$\FF=\e^{-\frac{\ii}{2}\theta^{a b}{X_a} \otimes X_b}~$, where the
$X_a$'s are arbitrary commuting vectorfields (and $\theta^{a b}$
is antisymmetric), then these three equations are equivalent.
They are satisfied if (and when  $\theta^{a b}$ is nondegenerate only if)
the vectorfields $X_a$ leave invariant the
Poisson structure (in particular this happens if they are
Hamiltonian vectorfields). 
The semiclassical limit of
equations~\eqn{compLATW1} and~\eqn{compLATW2} implies that the
Poisson structure $P$ associated with the twist $\FF$ is
compatible with the Poisson structure $\La$ on the manifold $M$.
Explicitly $[P,\La]=0$, where $[~,~]$ is the Schouten-Nijenhuis
bracket.

Condition~(\ref{compLATW1}) implies that
\eq
\{f,g\}_\st=\{\bar\ff^\al(f) , \bar\ff_\al(g)\}~, 
\en 
and that Hamiltonian vectorfields are undeformed,  
\be
X^\st_f:=\langle \La,
df\rangle_\st=\langle \La, df\rangle=X_f. \label{boh}
\ee
It can be proven that conditions~(\ref{compLATW1}) and~(\ref{compLATW2}) imply
the following compatibility between the twist and Hamiltonian vectorfields
\eq
\bar\ff^\al\otimes \bar\ff_\al(X_h) =\bar\ff^\al\otimes
X_{\bar\ff_\al(h)}~, \en
\eq
\bar\ff^\al(X_h)\otimes \bar\ff_\al=X_{\bar\ff^\al(h)}\otimes
\bar\ff_\al~. \en

The $\st$-Jacoby identity, that is equivalent to property
(\ref{closure}), easily follows from these equations because of
linearity
\eq
[X_f,X_g]_\st=[\bar\ff^\al(X_f),\bar\ff_\al(X_g)]=[X_{\bar\ff^\al
(f)},X_{\bar\ff_\al(g)}]= X_{\{\bar\ff^\al
(f),\bar\ff_\al(g)\}}=X_{\{f,g\}_\st}~. \en 
Because of this
property and of the Leibniz rule (\ref{LeibHam}) (or better the
coproduct rule (\ref{copHam})) we have that also for a general twist 
with a compatible Poisson bracket {\sl Hamiltonian vector fields are
a $\star$-Lie subalgebra of the $\star$-Lie algebra of vectorfields}.

\subsection{Time Evolution and Constants of Motion} 
The study of the noncommutative phase space geometry is 
here applied to briefly discuss time evolution and symmetries 
in deformed mechanics. We consider point particles  
on space with usual Moyal-Weyl noncommutativity given by the 
$\theta$-constant twist 
$\FF=\e^{-\frac{\ii}{2}\theta^{ij}\partial_i \otimes
\partial_j}$.

A natural definition of time evolution is
\be
\dot f=-\mathcal{L}_{X_H}^\star  f = -
\{H,f\}_\star\label{4}~.
\ee
As noticed in (\ref{boh}), we see that the time evolution generator 
$X_H = \frac{\del H}{\del x^i} \frac{\del}{\del p_i} -\frac{\del
H}{\del p_i} \frac{\del}{\del x^i}$
is the same as the undeformed one; it is its action $\mathcal{L}^\st$ 
on functions that is deformed. Indeed
in general $\{H,f\}_{\star}\not=\{H,f\}$ and therefore time 
evolution is different from the undeformed one.
Equation~\eqn{4} should be considered as an equation for 
the deformed pull-back flow 
$({\sigma_t^*})_\st$
(cf.(\ref{pbflow})), 
\eq
\frac{\rm d}{{\rm d} t}\,({\sigma_t^*})_\st=-({\sigma_t^*})_\st\circ 
{\mathcal L}^\st_{X_H}~.\label{pbflownc}
\en
Equation~\eqn{4}, (or ~\eqn{pbflownc}) can be formally integrated 
\be
({\sigma_t^*})_\st \,f= \exp{(-t \mathcal{L}_{X_H}^\star}) f = f -
t \mathcal{L}^\st_{X_H} f + \frac{1}{2} t^2
\mathcal{L}^\st_{X_H}(\mathcal{L}^\st_{X_H} f) + \ldots \label{66}~.
\ee
A more explicit expression of this formula is obtained if we denote by 
$\xi^a$ the phase space coordinates $x^i,p_j$, and if 
we correspondingly expand the Hamiltonian vectorfield as $X_H=X_H^a\,\partial_a$, where 
$\partial_a=\frac{\del}{\del \xi^a}$. Then we have  
\be
({\sigma_t^*})_\st \,f= \exp{(-t \mathcal{L}_{X_H}^\star}) f = f -
t X_H^a\st_{\,}\partial_a f + \frac{1}{2} t^2 X_H^a\st_{\,} \partial_a 
(X_H^a\st_{\,}\partial_a f)+ \ldots \label{7}
\ee
Another expression for $(\sigma_t^*)_\st$ is
$(\sigma_t^*)_\st={\mathcal L}^\st_{e_\st^{t X_H}}$ where
the $\st$-exponential ${e_\st^{t X_H}}$ is obtained with the $\st$-product in 
$U\Xi_\st$, and ${\mathcal L}^\st$ represents ${e_\st^{t X_H}}$ as a differential operator on functions.

It is easy to verify the one parameter group property
 $(\sigma_t^*)_\st\circ (\sigma_s^*)_\st =(\sigma_{t+s}^*)_\st$. On the other hand the deformed Leibniz rule for $\mathcal{L}^\st_{X_H}$ implies 
$({\sigma_t^*})_\st (f\st g)\ne ({\sigma_t^*})_\st f\st ({\sigma_t^*})_\st g$, 
as well as
\eq(\sigma_t^*)_\st f(x,p) \ne f(x(t),p(t))~,
\en where $x^i(t)=(\sigma_t^*)_\st x^i$, 
$p_j(t)=(\sigma_t^*)_\st p_j$. 
\sk
A constant of motion is a function $Q$ on phase space that satisfies
\eq
\{H,Q\}_\st=0~. \label{como}
\en
If 
\be
\{Q,H\}_\st=0 \label{symo}
\ee
we say that the Hamiltonian is  invariant under  
 the vectorfield $X_Q$ (because 
$\{Q,H\}_\st=\mathcal{L}^\st_{X_Q} H$).
Since the $\st$-Poisson bracket is not antisymmetric \eqn{como} and \eqn{symo}
 are independent equations.
  
Notice that for translation invariant Hamiltonians the 
time evolution equation as well as the notion of constant of motion 
are undeformed. Then \eqn{como} and \eqn{symo} coincide. 
Using the $\st$-Jacoby identity we have that 
the $\st$-bracket $\{Q,Q'\}_\st$ of two constants of motion is again a 
constant of motion. We conclude that the subspace of Hamiltonian vector 
fields $X_Q$ that $\st$-commute with $X_H$ form a $\st$-Lie subalgebra of 
the $\st$-Lie algebra of Hamiltonian vectorfields. The $\st$-symmetry 
algebra of constants of motion.

Examples of translation invariant Hamiltonians include all point particles 
Hamiltonians whose potential depends only on the relative distance of the 
point particles involved.  We also see that this formalism is quite well 
suited for field theory Hamiltonians that have potentials like 
$\int\!\dd^3\!x \,\overline\phi(x)\st\phi(x)\st\overline\phi(x)\st\phi(x)$ and
are translation invariant. 


\subsubsection{Example: {The Harmonic Oscillator} }
In this subsection we see our deformed point mechanics 
at work on a simple example that does not admit translation invariance. 

We consider the 
harmonic oscillator in two noncommutative space dimensions. We study its 
equation of motion, the constants of motion and the invariances of 
the Hamiltonian.
Angular momentum is not conserved, but a deformed version is.
Viceversa, a deformation of this oscillator conserves usual 
angular momentum.  

The results here presented are not used in the later sections on  field 
theory.

Let 
\be
H= \frac{1}{2}(x^i\star  x^j \delta_{ij} + p_i\star p_j
\delta^{ij})= \frac{1}{2}(x^i x^j \delta_{ij} + p_i p_j \delta^{ij})
\label{hamharm}
\ee
\be
L=\varepsilon_{i}^j x^i \st p_j=\varepsilon_{i}^j x^i  p_j~~~~~~~~~~~~~~~~~~~~~~~~~~~~~~~~~~~~~~~
\ee
be the Hamiltonian and the angular momentum of the 2-dimensional 
harmonic oscillator.

Since
\eq
\{h,f\}_{\star}=\{h,f\}\label{hfpar}
\en
if $h$ and $f$ are sums of functions that depend only on the coordinates
$x^i$ or the momenta $p_j$, we have the undeformed 
equations $\{H,H\}_\st=\{H,H\}=0$ and 
\bea
\dot x^i &=& -\{H, x^i\}_\star=-\{H, x^i\}~, \nonumber\\
\dot p_j &=& -\{H,p_j\}_\star=-\{H,p_j\} ~.\label{5}
\eea
On the other hand neither the angular momentum is a constant of motion
\be
\dot L= -\{H,L\}_\st=-\mathcal{L}^\st_{X_H} L=
-\frac{\ii}{2}\varepsilon_{ij}\theta^{ij}=-\ii\theta \label{8}
\ee
(we have defined 
$
\theta^{ij}= \theta \varepsilon^{ij}\,
$),
nor the Hamiltonian is rotation invariant, indeed we have
$\mathcal{L}^*_{X_L} H =\{L,H\}_\st=
-\ii\theta
$. 
From~\eqn{7} 
the time evolution of the angular momentum is 
$
(\sigma_t^*)_\star L= L -  \ii\theta t
$. 

We recall that the classical harmonic 
oscillator is a maximally superintegrable system, that is, it has 
$3~ (=2d-1)$ constants of motion which are functionally independent. 
For example we can consider 

\be 
H~~,~L~~,~K=(x^1)^2-(x^2)^2+ (p_1)^2-(p_2)^2~~,~
T=x^1 x^2 + p_1 p_2~.  \label{tconst}
\ee
Only three of the above constants of motion are functionally independent.
The third and fourth constants have the interesting property of being 
preserved in our twist-deformed setting. Indeed from (\ref{hfpar}) it 
immediately follows
\be
\{H,K\}_\star=\{K,H\}_\star=0~~,\{H,T\}_\star=\{T,H\}_\star=0~.
\ee
Therefore the $\st$-harmonic oscillator 
remains a superintegrable system, but loses rotational invariance.
\sk
Deformations $L_\st$ of the angular momentum $L$ can however be constants of 
motion. For example we have the two functionally independent deformations
\beqa 
L_\st'&=& L - \ii\theta  {\rm{Arctan}}\Big(\frac{x^1}{p_1}\Big)~,\nn\\
L_\st''&=& L - \ii\theta  {\rm{Arctan}}\Big(\frac{x^2}{p_2}\Big)~.\label{Ldefo}
\eeqa
that satisfy $\{H,L_\st'\}_\st=0$ , $\{H,L_\st''\}_\st=0$.
In order to prove this statement it is instructive to consider an
arbitrary $\theta$-deformation of $L$,
\be
L_\st=\sum_{n=0}^\infty \theta^n L_n~,
\ee
where $L_0=L$ and all coefficients $L_n$ are $\theta$-independent functions on phase space. We determine these coefficients by requiring $L_\st$ to be a 
constant of motion,
\eq
\{H, L_\st\}_\star=\sum_{n=0}^\infty\theta^n\{H,L_n\}_\star=0~.
\en
Since for any function $f$ on phase space we have
\be
\{H,f\}_\star=\{H,f\} -\frac{\ii}{2} \theta 
\varepsilon^i_j \frac{\del}{\del x^i}\frac{\del}{\del p_j} f~,
\en
$L_\st$ is a constant of motion if
\eq
\sum_{n=0}^\infty \theta^n \{H,L_n\} -\frac{\ii}{2}
\theta^{n+1} \varepsilon^i_{j}\frac{\del}{\del x^i}\frac{\del}{\del p_j} L_n=0~.
\ee
All the coefficients in this $\theta$-expansion have to vanish 
and we then obtain the recursive relation
\eq
\{H,L_{n+1}\}= \frac{\ii}{2}\varepsilon^i_{j} \frac{\del}{\del x^i}
\frac{\del}{\del p_j} L_n \label{newconst}
\en
with the initial condition $L_0=L$.
At first order in $\theta$ we have
$
\{H,L_1\}= 
\ii\,,
$ that is
\be
\Big(x^j\frac{\del}{\del p_j}-p_j\frac{\del}{\del x^j}\Big) L_1=\ii  
\label{pde}~.
\ee
Since the left hand side preserves the degree of any homogeneous 
polynomial in the coordinates $x^i$ and $p_j$, no analytic function 
on phase space can solve this equation. If we relax the analyticity 
condition we 
find two independent solutions
\beqa
L_1'&=& -\ii {\rm{Arctan}}\Big(\frac{x^1}{p_1}\Big) \label{Ldefo1}~,\\ 
L_1''&=& -\ii {\rm{Arctan}}\Big(\frac{x^2}{p_2}\Big)  \label{Ldefo2}~.
\eeqa
In order to solve (\ref{newconst}) we can choose all higher order coefficients 
$L_n$ with $n\geq 2$ to be zero. We thus obtain the two solutions
(\ref{Ldefo}). 
Notice that, unlike $H$ and $T$, the constants of motion 
(\ref{Ldefo}) do not $\star$-commute with themselves.

As an instance of our general comment on the independence of \eqn{como} 
and \eqn{symo}, that is to say, on the independence of the notions of 
constant of motion and invariance, 
we observe that the two constants of motion 
\eqn{Ldefo} do not generate symmetries of the Hamiltonian. 
It can be easily verified that solutions of \eqn{symo} are given instead 
by the complex conjugates of \eqn{Ldefo}. 

\sk
We find also interesting to study deformations $H_\st$ of the harmonic 
oscillator Hamiltonian that admit the undeformed angular momentum $L$ as 
constant of motion. The aim, like in \cite{Lorek}, is to consider new 
dynamical systems that 
may be highly nontrivial if thought in commutative space 
(the equation of motion ~\eqn{4} or ~\eqn{5} can just be seen as a partial 
differential equation on commutative spacetime)
but that analyzed in the noncommutative Hamiltonian mechanics 
framework show the same constants of motion, and possibly richness of 
symmetries and integrability, as the undeformed ones.

We therefore consider the power series 
\eq
H_\st=\sum_{n=0}^\infty \theta^n H_n
\en   
with $H_0=H$, and determine the coefficients $H_n$ (that are functions on 
phase space) by requiring
$
\{H_\st,L\}_\star=\sum_{n=0}^\infty \theta^n \{H_n,L\}_\star=0~.
$
Since
\eq
\{f,L\}_\star=\{f,L\}-\frac{\ii}{2} \theta 
 (\del_1^2+\del_2^2) f~,
\en
by setting $f=H_\st$ we obtain the recursion relation
\be
\{H_{n+1},L\}=\frac{\ii}{2}  (\del_1^2+\del_2^2) H_n \label{eqH}~,
\ee
and in particular 
\be
\{H_{1},L\}=\frac{\ii}{2}  (\del_1^2+\del_2^2) H_0 = \ii~. \label{firco}
\ee
This yields a partial differential equation similar to \eqn{pde}
\be
\varepsilon_j^k\Big (x^j\frac{\del}{\del x^k}- p_k\frac{\del}{\del p_j}
\Big)H_1=\ii  
\label{pde2}
\ee
As in the previous calculation since the operator on the left hand side 
preserves the degree of a homogeneous polynomial in $x^i$ and $p_j$, no 
analitic solution is possible.
Comparison with \eqn{pde} however gives the solutions,
\beqa
H_1^{'}&=& -\ii {\rm{Arctan}}\Big(\frac{p_1}{p_2}\Big) \label{Hdefo1}\\ 
H_1^{''}&=& -\ii {\rm{Arctan}}\Big(\frac{x^1}{x^2}\Big)  \label{Hdefo2}
\eeqa
Again it can be checked that all the subsequent equations in \eqn{eqH} are 
satisfied with the choice $H_i=0, i\ge 2$, therefore we have two possible 
deformations of the Hamiltonian which admit the angular momentum as a constant 
of  motion
\beqa 
H_\st'&=& H - \ii\theta  {\rm{Arctan}}\Big(\frac{p_1}{p_2}\Big)\nn\\
H_\st''&=& H - \ii\theta  {\rm{Arctan}}\Big(\frac{x^1}{x^2}\Big). \label{Hdefo}
\eeqa
Notice however that 
$\{L, H_\st'\}_\st \ne 0$ and  
$\{L, H_\st''\}_\st \ne 0$, that is, \eqn{Hdefo} are not invariant 
under rotations.
Rotational invariance is fulfilled if we consider the complex conjugates of 
\eqn{Hdefo}.
   
It is interesting to note that, unlike  the deformations of the angular 
momentum \eqn{Ldefo}, both the deformations \eqn{Hdefo} 
$\star$-commute with themselves.
The first Hamiltonian $H_\st'$ is nonlocal, while the second one is local.
They are both real if we consider the parameter $\theta$ to be purely 
imaginary.
We will not deepen their analysis here because it exulates form 
the scopes of the present article.

\section{Classical Field Theory}
\setcounter{equation}{0}
We generalize the twist setting to the case of an infinite number of degrees of freedom.
In this case the position and momenta
generalize to the fields $\Phi(x)$ and $\Pi(x)$ with $x\in\mathbb
R^d$ ($\real^{d+1}$ being spacetime). The algebra $\mathsf{A} $ is
an algebra of functionals, it is the algebra of functions on $N$
where in turn $N$ is the function space:
\be
N={\mbox{Maps}}\,(\mathbb R^d\rightarrow\mathbb R^2)
\label{spacefields}~.
\ee
Here we are considering a scalar field theory, in a more general
case $\mathbb R^2$ (with its coordinates $\Phi$ and $\Pi$) is
substituted by the proper target space. The generalization to
$\mathbb R^{2s}$ (with $s$ scalar fields) is immediate. Particle
mechanics with phase space $\real^{2d}$ is recovered by
considering that $\mathbb R^d$ in~\eqn{spacefields} collapses to
$d$ points. \sk We define the Poisson bracket between the
functionals $F,G\in\mathsf{A} $ to be
\be
\{F,G\}=\int\dd^{d} x \,\,\,\frac{\delta F}{\delta
\Phi}\frac{\delta G}{\delta \Pi}  - \frac{\delta F}{\delta
\Pi}\frac{\delta G}{\delta \Phi}
\ee
The fields $\Phi(x)$ and $\Pi(x)$ for fixed $x$ can be considered
themselves a family of functionals parametrized by $x\in\real^n$,
for fixed $x$, $\Phi(x)$ is the functional that associates to
$\Phi$ and $\Pi$ the value $\Phi(x)$; similarly with $\Pi(x)$).
Their brackets are\footnote{In order to avoid considering
distributions we should work with smeared fields $
\Phi(f)=\int\dd^{d} \!x\,\, f(x) \Phi(x)$ and $\Pi(g)=\int\dd^{d}
\!x \, \,g(x) \Pi(x)$. The smeared version of the Poisson bracket
is then $ ~\{\Phi(f),\Pi(g)\}=\int\dd^{d} x f(x) g(x)~. $}
\bea
\{\Phi(x),\Phi(y)\}&=&0~,\nonumber\\
\{\Pi(x),\Pi(y)\}&=&0~,\nonumber\\
\{\Phi(x),\Pi(y)\}&=&\delta(x-y)~.
\eea

Now let space $\mathbb R^d$ become the noncommutative Moyal space.
The algebra of functions on $\mathbb R^d$  and the algebra
\eqn{spacefields} become noncommutative with noncommutativity
given by the twist (\ref{MWTW}), $\FF=\e^{-\frac{\ii}{
2}\theta^{ij}\frac{\partial}{\partial x^i} 
\otimes\frac{\partial}{\partial x^j}}\,.$

The twist lifts to the algebra $\mathsf{A}$ of functionals \cite{Wessgauge} 
so that this latter too becomes noncommutative. This is
achieved by lifting to $\mathsf A$ the action of 
infinitesimal translations.
Explicitly $\frac{\partial}{\partial x^i}$
is lifted to $\del^*_i$ acting on $\mathsf{A}$ as,
\be
\del_i^* G:=-\int\dd^d x \,\,\del_i\Phi(x)\frac{\delta G}{\delta\Phi(x)}
+\del_i\Pi(x)\frac{\delta G}{\delta\Pi(x)} 
\label{funder} ~.
\ee
Therefore on functionals the twist is represented as
\be
\mathcal F=\e^{-\frac{\ii}{2}\theta^{ij}\!\int\!\dd^{\!d}\! x
\left(\del_i\Phi\frac{\delta}{\delta\Phi(x)}+\del_i\Pi\frac{\delta}{\delta
\Pi(x)}\right)\,\otimes\, \int\!\dd^{\!d}\!
y\left(\del_j\Phi\frac{\delta}{\delta
\Phi(y)}+\del_j\Pi\frac{\delta}{\delta\Pi(y)}\right)}
\label{lifttwist}~.
\ee
The associated $\st$-product is
\eq
F\st G= \bar\ff^\alpha(F)\bar\ff_\alpha(G)~.\label{Funstar} 
\en
We can regard $\Phi(x)$ as the functional 
$\Phi(x)=\int \dd^dz \,\delta(x-z) \,\Phi(z)$ 
that associates to the function $\Phi$ its value in $x$.
In particular we can consider the $\st$-product between functionals  $\Phi(x)\st\Phi(y)$.
If $x=y$ then $\Phi(x)\st\Phi(y)=(\Phi\st\Phi)(x)$ where this latter 
$\st$-product is the usual one with the {\sl function} $\Phi$.
\sk
\noi{\sl Note 3. The twist
$\FF=\e^{\frac{\ii}{2}\theta^{ij}\del_i\otimes\del_j}$ gives rise to  the
$\st$-Lie algebra of infinitesimal diffeomorphisms of Subsection
2.2; similarly the twist~(\ref{lifttwist}) yields  the $\st$-Lie
algebra of infinitesimal functional variations. The former
$\st$-Lie algebra is generated by the $\st$-Lie derivatives along
vectorfields ${\cal L}^\st_u$, the latter $\st$-Lie algebra is
generated by the $\st$-functional variations $\delta^\st_\epsi$.
We briefly discuss this $\st$-Lie algebra in the appendix.}


\sk
Let us consider the canonical Poisson tensor
\eq
\Lambda=\int \dd^d x  \left( \frac\delta{\delta\Phi(x)} \otimes
\frac\delta{\delta\Pi(x)} - \frac\delta{\delta\Pi(x)} \otimes
\frac\delta{\delta\Phi(x)}\right) 
\en 
and verify that it is
compatible with the twist~(\ref{lifttwist}), i.e. that
relations~(\ref{compatibility})-(\ref{compLATW2})  hold. We unify the phase space
coordinates notation by setting
\eq\label{onenot}
\Psi^a=(\Phi, \Pi)~. \en Then the action of 
infinitesimal translations on functionals is rewritten as
\eq\label{4.9}
\del_i^* =-\int\dd^d y
\,\,\del_{y^i}\Psi^a(y)\,\frac{\delta}{\delta\Psi^a(y)}~. \en We
compute
$$\del_i^*\Big(\int \dd^d x \,\frac{\delta}{\delta\Psi^b(x)}
          \otimes  \frac{\delta}{\delta\Psi^c(x)}\Big)=
\int \dd^d x\, \del_i^*\Big(\frac{\delta}{\delta\Psi^b(x)}\Big)
          \otimes  \frac{\delta}{\delta\Psi^c(x)}+
\frac{\delta}{\delta\Psi^b(x)}
          \otimes  \del_i^*\Big(\frac{\delta}{\delta\Psi^c(x)}\Big)$$
\eqa
&&\nn\\[-1.2em]
&&~~~~~~~~ =\,\int \dd^d x
\,\Big[\del_i^*\,,\,\frac{\delta}{\delta\Psi^b(x)}\Big]
          \otimes  \frac{\delta}{\delta\Psi^c(x)}+
\frac{\delta}{\delta\Psi^b(x)}
          \otimes  \Big[\del_i^*\,,\,\frac{\delta}{\delta\Psi^c(x)}\Big]\nn\\[.4em]
&&~~~~~~~~ =\int \dd^d x \, \dd^dy \,\,\del_{y^i}\delta(x-y)\,
\frac{\delta}{\delta\Psi^b(y)}
          \otimes  \frac{\delta}{\delta\Psi^c(x)}\,+\,
\frac{\delta}{\delta\Psi^b(x)}
          \otimes  \del_{y^i}\delta(x-y)\frac{\delta}{\delta\Psi^c(y)}\nn\\[.4em]
&&~~~~~~~~=\,0
\ena
where in the last equality we have exchanged the dummy $x$ and $y$
variables of the second addend, and used that
$\del_{y^i}\delta(x-y)=-\del_{x^i}\delta(x-y)$. The vanishing of
this expression implies the compatibility relations (\ref{compatibility})-(\ref{compLATW2}).

The compatibility between the Poisson tensor and the twist assures that we 
have a well defined notion of
deformed Poisson bracket, $\{~,~\}_\st~:~\mathsf{A}\otimes
\mathsf{A}\rightarrow \mathsf{A}$,
\eq
\{F,G\}_\star:= \{\bar\ff^\alpha(F),\bar\ff_\alpha(G)\}
\label{funpoi}~. \en 
This bracket satisfies
\eqa
\{F,G\}_{\star}&=&-\{\bar\rr^\alpha(G),\bar
\rr_\alpha(F)\}_{\star}\label{stantisF}\\
\{F,G\star H\}_\star &=&\{F,G\}_\star \star H +\bar
\rr^\alpha(G)\star \{\bar \rr_\alpha(F),H\}_\star
\label{leibtwistF}\\
\{F,\{G, H\}_\star\}_\star &=&  \{\{F,G\}_\star,H\}_\star +
\{\bar\rr^\alpha(G), \{\bar\rr_\alpha(F),H\}_\star\}_\star
\label{jacobitwistF}
\ena
In particular the $\st$-brackets among the fields are undeformed
\eqa
&&\{\Phi(x),\Pi(y)\}_\star=\{\Phi(x),\Pi(y)\}=
\delta(x-y)\label{PBfipi}~,\\[.4em]
&&\{\Phi(x),\Phi(y)\}_\star=\{\Phi(x),\Phi(y)\}=0~,\\[.4em]
&&\{\Pi(x),\Pi(y)\}_\star=\{\Pi(x),\Pi(y)\}=0~.
\ena
We prove the first relation 
\bea
\{\Phi(x),\Pi(y)\}_\star&=&
\{\bar\ff^\alpha(\Phi(x)),\bar\ff_\alpha(\Pi(y))\}\nonumber\\
&=&\{\Phi(x),\Pi(y)\}-\frac\ii 2 \theta^{ij}{\textstyle\left\{\int\!\dd^d
z\, \del_i\Phi(z)\delta(x-z),\int\!\dd^d
w\,\del_j\Pi(w)\delta(y-w)\right\}}+ O(\theta^2) \nonumber\\
&=&\{\Phi(x),\Pi(y)\}-\frac\ii 2
\theta^{ij}\del_{y^j}\del_{x^i}\delta(x-y) + O(\theta^2)
\nonumber\\
&=&\{\Phi(x),\Pi(y)\}~;
\label{4.18}\eea
the second term in the third line vanishes
because of symmetry, as well as higher terms in $\theta^{ij}$.

We conclude that for Moyal-Weyl
deformations also in the field theoretical case the $\st$-Poisson
bracket just among coordinates is unchanged. It is however important to stress 
that this is not the case in general. For nontrivial functionals of the 
fields we have
\be
\{F,G\}_\st\ne\{F,G\}   \label{nt}~.
\ee

We now expand $\Phi$ and $\Pi$ in Fourier modes:
\bea
\Phi(x)&=&\int\frac{\dd^{d} k}{(2\pi)^{d}\,\sqrt{2E_k}}
\left(a(k)\,\e^{\ii kx} +a^*(k)\e^{-{\ii}kx}\right)\nonumber\\
\Pi(x)&=&\int\frac{\dd^{d} k}{(2\pi)^{d}}
(-\ii\hbar)\sqrt{\frac{E_k}{2}} \left(a(k) \e^{{\ii} kx} -
a^*(k)\e^{-\ii kx}\right) \label{expphipi}
\eea
where $E_k=\sqrt{m^2+{\vec p}^{\,2}}=\sqrt{m^2+\hbar^2{\vec
k}^{\,2}}$, and $kx=\vec k \cdot \vec x =\sum_{i=1}^{d}k^ix^i$.
We use the usual undeformed Fourier decomposition because indeed
are the usual exponentials that, once we also add the time
dependence part, solve the free field equation of motion on
noncommutative space $(\hbar^2\del^\mu\del_\mu+m^2)\Phi=0$. This
equation is the same as the one on commutative space because the
$\st$-product enters only the interaction terms.

The expressions of the fields $\Phi$ and $\Pi$ in terms of the
Fourier coefficients $a$ and of their complex conjugate $a^*$ can
be inverted to give:
\eqa
a(k)&=&\int\!\dd^d x \Big(\sqrt{\frac{E_k}{2}}\,\Phi(x)+\frac{\ii}
\hbar{\sqrt{\frac{1}{2E_k}}}\,\Pi(x)\Big)\,\e^{-\ii kx}\nonumber\\[.3em]
a^*(k)&=&\int\!\dd^d x
\Big(\sqrt{\frac{E_k}{2}}\,\Phi(x)-\frac{\ii}{\hbar}{\sqrt{\frac{1}{2E_k}}}\,
\Pi(x)\Big)\,\e^{{\ii}kx}
\label {inverrel}
\ena
From these formulae we see that for each value of $k$, $a(k)$ and
$a^*(k)$ are functionals of $\Phi$ and $\Pi$. We therefore can
consider the $\st$-product between these functionals
as defined in (\ref{Funstar}). In order to explicitly calculate
the $\st$-product we observe that the action (\ref{funder}) of the infinitesimal translations $\frac\del{\del x^i}$ on 
the functionals $a$ and $a^*$ (that for ease of notation we here just denote by $\del_i$) is 
\eq
\partial_i a(k)= -\ii k^i a(k)=\ii k_i a(k)~~,~~~~
\partial_i a^*(k)= \ii k^i a(k)= -ik_i a^*(k)~~. \label{dela}
\en 
We find instructive to write the $\st$-product in few simple
cases
\eqa
a(k)\st a(k')
&=&\e^{-\frac{\ii}{2}\theta^{ij}\,k_ik'_j}a(k)a(k') ~~~~~,~~~~~~
a^*(k)\st a^*(k')
=\e^{-\frac{\ii}{2}\theta^{ij}k_ik'_j}\,a^*(k)a^*(k')~~, \nonumber\\[1em]
a^*(k)\st a(k')
&=&\e^{\frac{\ii}{2}\theta^{ij}k_ik'_j}\,a^*(k)a(k') ~~~~~\,,~~~~~~~
a(k)\st a^*(k')
=\e^{\frac{\ii}{2}\theta^{ij}k_ik'_j}\,a(k)a^*(k')~~, \nn
\ena
and more in general
$$
a(k^{(1)})\st a(k^{(2)})\st\ldots a(k^{(m)})= \e^{-\frac{\ii}{2}
\theta^{ij}\sum_{r<s}k^{(r)}_ik^{(s)}_j}
a(k^{(1)})\,a(k^{(2)})\,\ldots a(k^{(m)})
$$
where $r,s=1,2\ldots m$. A similar formula holds for mixed $a$ and
$a^*$ products.

We finally easily calculate the Poisson bracket 
among the Fourier modes using the definition  (\ref{funpoi}) and the 
functional expressions of $a(k)$,
$a^*(k)$ in terms of $\Phi$ and $\Pi$ (\ref{inverrel}), or equivalently from (\ref{funpoi}) and (\ref{dela}). We obtain
\be
\{a(k),a^*(k')\}_\st=e^{\frac{\ii}{2}\theta^{ij}k_i k'_j} \{a(k),a^*(k')\}=
-\frac\ii\hbar (2\pi)^d\delta(k-k')~,\label{poires}
\ee
where we used the undeformed relation 
$\{a(k),a^*(k')\}
 =-\frac\ii\hbar (2\pi)^d\delta(k-k')$.
The phase drops out in \eqn{poires} because the delta contributes 
only for $k=k'$, in which case the antisymmetry of $\theta$ 
forces the exponent to be zero. 
We similarly have
\be
\{a(k),a(k')\}_\st=0~~~~,~~~~~~~ \{a^*(k),a^*(k')\}_\st=0~~. \en
As for our comment related to \eqn{nt}, this is a good place to check 
nontriviality 
of the twisted Poisson bracket. Although it is equal to the untwisted one 
for linear combinations of the Fourier modes, it is easily verified that 
it yields a different result, 
involving nontrivial fases, as soon as we consider Poisson brackets of powers 
of $a$, $a^*$.  

\section{Field Quantization}
\setcounter{equation}{0}

We now formulate the canonical quantization of scalar fields on
noncommutative space.  Associated to the algebra $\mathsf{A}$ of 
functionals $G[\Phi,\Pi]$ there is the algebra $\widehat{\mathsf{A}}$
of functionals $\hat G[\hat\Phi,\hat\Pi]$ on operator valued fields. 
We lift the twist to $\widehat{\mathsf{A}}$ and then deform this algebra 
to $\widehat{\mathsf{A}}_\st$
by implementing once more 
the twist deformation principle (\ref{generalpres}).
We denote by $\hat{\partial}_i$ the lift to $\widehat{\mathsf{A}}$ 
of $\frac\partial{\partial x^i}$; 
for all $\hat G\in \widehat{\mathsf{A}}$, 
\be
\hat\del_i \hat G:=- \int\dd^d x \,\,\del_i\hat\Phi(x)\frac{\delta \hat G}{\delta\hat\Phi(x)}
+\del_i\hat\Pi(x)\frac{\delta \hat G}{\delta\hat\Pi(x)} 
\label{dhat}~;
\ee
here $\del_i\hat\Phi(x)\frac{\delta \hat G}{\delta\hat\Phi(x)}$ stands for 
$\int\dd^d\ell\,\del_i\Phi_\ell(x)\frac{\delta \hat G}{\delta\Phi_\ell(x)}$, where like in (\ref{expphipi}) we 
have expanded the operator $\hat\Phi(x)$ 
as $\int\dd^d \ell \,\Phi_\ell(x)\hat{\sf a}(\ell)$ (and similarly for $\hat\Pi(x)$). 

Consequently the twist on operator valued functionals reads
\be
\hat{\mathcal F}=\e^{-\frac{\ii}{2}\theta^{ij}\!\int\!\dd^{\!d}\! x
\left(\del_i\hat\Phi\frac{\delta}{\delta\hat\Phi(x)}+\del_i\hat\Pi
\frac{\delta}{\delta
\hat\Pi(x)}\right)\,\otimes\, \int\!\dd^{\!d}\!
y\left(\del_j\hat\Phi\frac{\delta}{\delta
\hat\Phi(y)}+\del_j\hat\Pi\frac{\delta}{\delta\hat\Pi(y)}\right)}
\label{lifttwist2}~.
\ee
In $\widehat{\mathsf{A}}_\st$ there is a natural notion 
of $\st$-commutator, according to 
the general prescription \eqn{generalpres}
\be\label{bobobo}
[~,~]_\st= [~,~]\circ {\mathcal F}^{-1}~.
\ee
This $\st$-commutator is $\st$-antisymmetric, is a $\st$-derivation in 
${\hat{\mathsf{A}}}_\st$ and satisfies the $\st$-Jacoby identity
\eqa
[\hat F,\hat G]_{\st} &= &-[\bar\rr^\alpha(\hat G),
\bar\rr_\alpha(\hat F)]_{\st} \label{stantisFc}\\
{}[\hat F,\hat G\st \hat H]_\st &=& [\hat F,\hat G]_\st \st \hat H +
\bar\rr^\alpha(\hat G)\st [\bar\rr_\alpha(\hat F),\hat H]_\star 
\label{leibtwistFc}\\
{}[\hat F,[\hat G, \hat H]_\star]_\star &=&  [[\hat F,\hat G]_\star,\hat H]_\star +
[\bar\rr^\alpha(\hat G), [\bar\rr_\alpha(\hat F),\hat H]_\star]_\star
\label{jacobitwistFc}
\ena
Finally, recalling the definition of the $\RR$-matrix it can be easily 
verified that  
\be\label{stconast}
[\hat F,\hat G]_\st= \hat F\st \hat G - 
\bar R^\alpha(\hat G)\st \bar R_\alpha(\hat F)
\ee
which is indeed the $\st$-commutator in  $\hat {\mathsf{A}}_\st$.
This $\st$-commutator \eqn{bobobo} has been considered in \cite{Zahn} 
(and was introduced in \cite{Bahns}). 
\sk
We studied four algebras and  brackets: 
 $({\mathsf{A}},\,\{~,~\})\,,~
(\widehat{\mathsf{A}},\,[~,~])\,,~ 
({\mathsf{A}_\st},\,\{~,~\}_\st)\,,~
(\widehat{\mathsf{A}}_\st , \,[~,~]_\st)~.$ Canonical quantization on 
noncommutative space is the map $\hbar_\st$ in the diagram

\eq
\begin{CD}
{\mathsf{A}}                   @>\hbar\,\,>>  {\widehat{\mathsf{A}}}\\
 @V \FF VV                                      @V \widehat{\FF} VV\\
{~\mathsf{A}_\st}@>\hbar_\st >>{~\widehat{\mathsf{A}}_\st}
\end{CD}
\en
\sk
\noi We define canonical quantization 
on nocommutative space by requiring 
this diagram to be commutative. Notice that the vertical maps, that with abuse of notation we have 
called $\FF$ and $\hat{\FF}$, are the identity map, indeed $\mathsf{A}=\mathsf{A}_\st$ and $\widehat{\mathsf{A}}=\widehat{\mathsf{A}}_\st$ as 
vectorspaces. Therefore we have 
$\hbar_\st=\hbar$. 
The map $\hbar_\st$ satisfies a $\st$-correspondence 
principle because 
$\st$-Poisson brackets go into $\st$-commutators
at leading order in $\hbar$ 
\[
~~~~~~
\begin{CD}
{\{F,G\}~}                   @>\hbar~\,>>  {{-\frac{\ii}{\hbar}[\hat F,\hat G]~~}}\\[.3em]
 @V \FF VV                                      @V \widehat{\FF} VV\\[.2em]
\end{CD}
\]
\eq\label{diag}
\boldsymbol{
\begin{CD} ~~~~~{\{F,G\}_\st\!\!}\ @>\pmb{\hbar}_\st >>
{{-\frac{\ii}{\pmb{\hbar}}[\hat F,\hat G]_\st}}
\end{CD}}
\en
\sk
\noi Indeed recall the definitions of the $\st$-Poisson bracket and of the 
$\st$-commutator and compute
\eq
\{F,G\}_\st=\{\bar\ff^{\alpha}(F),\bar\ff_{\alpha}(G)\}\,\stackrel{\hbar}{\longrightarrow}\,-\frac{\ii}{\hbar} 
[\,\widehat{\bar\ff^{\alpha}( F)}\,,\,\widehat{\bar\ff_{\alpha}( G)}\,]
=-\frac{\ii}{\hbar} [\bar\ff^{\alpha}(\hat F),\bar\ff_{\alpha}(\hat G)]
=-\frac{\ii}{\hbar} [\hat F,\hat G]_{\st_{}}\label{5.9}
\en
The second equality holds because the lifts 
(\ref{funder}) and (\ref{dhat}) 
of $\frac\del{\del x^i}$ satisfy
\eq
\widehat{\del^*_iG}=\hat{\del}_i\hat G~,
\en
(as is most easily seen from (\ref{dela}) and (\ref{delonp})). 
\sk
From \eqn{bobobo}, repeating the passages of \eqn{4.18} we obtain  (in accordance with \eqn{5.9}) the $\st$-commutator of the fields $\hat \Phi$ and $\hat \Pi$,
\be
[\hat \Phi(x), \hat\Pi(y)]_\st = i\hbar\delta(x-y)~. \label{fipicom}
\ee
As a further confirmation that our quantization map $\hbar_\st=\hbar=\widehat{~{~}{~}}\,$
implements the $\st$-correspondence principle between
$\st$-Poisson brackets and $\st$-commutators we notice that to all orders in $\hbar$
\be
\widehat{\{\Phi,\Pi\}_\st}=\frac{\ii}{\hbar} [\hat \Phi,\hat \Pi]_\st~.
\label{stcomm} 
\en 
\sk
Concerning the creation and annihilation operators, they are functionals 
of the operators $\hat\Phi$, $\hat\Pi$ through the quantum analogue 
of the classical functional relation \eqn{inverrel}. Using \eqn{dhat}
we have (cf. \eqn{dela})
\be
\del_i \hat a(k)=  \ii k_i \hat a(k)~~,~~~~\del_i \hat{a}^\dagger(k)= - \ii k_i \hat{a}^\dagger(k) \label{delonp}~,
\ee
where here for ease of notation we have just denoted the lift of 
the infinitesimal translations $\frac\del{\del x^i}$ by $\del_i$.
Their $\st$-commutator follows from \eqn{fipicom} and the quantum analogue of \eqn{inverrel} (or from \eqn{bobobo} and \eqn{delonp}, or also
from \eqn{poires} and linearity of \eqn{stcomm}), 
\be
[\hat a(k),\hat a^\dagger(k')]_\star=(2\pi)^d\delta(k-k')~.\label{ppp}
\ee

In order to compare this expression with similar ones which have been found
in the literature \cite{FioreSchupp,bal1,bal2,LVV,Bu,kulish,fiorewess} 
it is useful to recall \eqn{stconast} and realize the action of the 
$\RR$-matrix. Since $\mathcal R=\mathcal F^{-2}$ we obtain that (\ref{ppp})
is equivalent to
\be
\hat a(k)\star \hat
a^\dagger(k')- \e^{-\ii\theta^{ij}k'_i k_j} \hat
a^\dagger(k')\star \hat a(k)= (2\pi)^d \delta(k-k') \label{acomm}~.
\ee
This relation first appeared in \cite{Grosse}. In the noncommutative QFT 
context it appears in \cite{kulish}, \cite{Bu}, and implicitly in \cite{Zahn} 
(it is also contemplated in \cite{fiorewess} as a second option). On the other 
hand
\cite{bal1,bal2,LVV,fiorewess}, starting from a different definition of 
$\st$-commutator, $[A \stackrel{\star}{,} B]:= A\star B- B\star A$, obtain 
deformed commutation relations 
of the kind $a_k a_{k'}^\dag - \e^{-\frac{\ii}{2} \theta^{ij}
k_ik'_j} a_k^\dag a_{k'} = (2\pi)^d\delta(k-k') $. These are different from 
(\ref{acomm}), indeed if we expand also 
the $\st$-product in (\ref{acomm}) we obtain the usual commutation relations
$\hat a(k)\hat
a^\dagger(k')-  \hat
a^\dagger(k')\hat a(k)= (2\pi)^d \delta(k-k')$.

As in the case of the $\st$-Poisson bracket, we have found that 
the $\st$-commutator of coordinate fields \eqn{fipicom}, and 
of creation and annihilation operators \eqn{ppp}, are equal to 
the usual undeformed ones.
Once again, we warn the reader that this is not true anymore for 
more complicated functionals of the coordinate fields,  in general
$[\hat F,\hat G]_\st\not=[\hat F, \hat G]$.


\setcounter{equation}{0}

\setcounter{section}{0}

\appendix{$\st$-Lie Algebra of Functional Variations}

In (\ref{sectwist}), we remarked that the twist
$\FF=\e^{\frac{-\ii}{2}\theta^{ij}\del_i\otimes\del_j}$ is an
element of the tensor product of $U\Xi$ by itself, the universal
enveloping algebra of the Lie algebra $\Xi$ of vectorfields.
Similarly the lifted twist~(\ref{lifttwist}) is an element of the
universal enveloping algebra $U\Upsilon$ of the Lie algebra
$\Upsilon$ of infinitesimal functional variations (on phase
space). In order to fully understand the lift~(\ref{lifttwist}) we
have to clarify the way the Lie algebra of infinitesimal
diffeomorphisms is a subalgebra of the Lie algebra of
infinitesimal functional variations.

Undeformed infinitesimal functional variations $\delta_\epsi$ are defined by
\eq
\delta_\epsi G:=\int \dd^d x \,\,\epsi^a(x)\frac\delta{\delta\Psi^a(x)}G
\en where $\Psi^a=(\Phi,\Pi)$ (more in general $\Psi^a$ are target
space coordinates), and where $\epsi^a(x)$ themselves can be
functionals.

Consider the map between vectorfields and infinitesimal functional variations
(with slight abuse of notation we denote this map by the symbol $\delta$)
\eqa
\delta~:~~ \Xi&\rightarrow & \Upsilon\nn\\
        u&\mapsto     & \delta_u\\
& &\delta_u G:=-\int \dd^d x \,\,u(\Psi^a)(x)\frac\delta{\delta\Psi^a(x)}G~.
\ena
This map is a Lie algebra map,
\eq
 \delta_{[u,v]}=[\delta_u,\delta_v] ~.
\en
If $u=\frac{\del}{\del x^i}$ then $\delta_u$ is just the lifted partial derivative 
$\partial^\st_i$ defined in (\ref{4.9}).
In order to proceed in the construction of the $\st$-Lie algebra
of functional variations we define $\st$-functional variations. According to
(\ref{generalpres}),
\eq\label{stfunvar}
\delta^\st_\epsi(G):=\bar\ff^\al(\delta_\epsi)\,(\bar\ff_\al(G))~;
\en 
where the action of $\bar\ff^\al$ on $\delta_\epsi$ is the adjoint
action in $U\Upsilon$,
$
\delta_\sigma(\delta_\epsi)=[\delta_\sigma,\delta_\epsi]\,,~(\delta_{\sigma_1}\delta_{\sigma_{2}})(\delta_\epsi)
=[\delta_{\sigma_1},
[\delta_{\sigma_{2}},\delta_\epsi]]$, and similarly
for higher products of variations $\delta_{\sigma_i}$.

The functional variation  $\delta_\epsi^\st$ satisfies the
Leibniz rule (cf. (\ref{defL}), (\ref{copHam}))
\eq\label{leibfunvar}
\delta^\st_\varepsilon (F\,\st\, G)=\delta_\varepsilon(F)\,\st\,
G+ \oR^\al(F)\,\st\, \big(\oR_\al(\delta_\epsi)\big)^* (G) \en
where $\oR_\al(\delta_\epsi)$ is itself a functional variation,
say $\delta_\sigma$, and
$\big(\oR_\al(\delta_\epsi)\big)^\st=\delta_\sigma^\st \,$. The
Leibniz rule is consistent (and actually follows) from the
coproduct rule
\eq\label{copleibfunvar}
\delta_\epsi\,\mapsto\, \Delta_\st(\delta_\epsi)=
\delta_\epsi\otimes 1 +\oR^\al\otimes \oR_\al(\delta_\epsi)~. \en
Finally also the formulae in this appendix hold for the most
generic twist; just replace $\oR^\al$ with $\ff^\beta (\oR^\al)\,
\ff_\beta$ in (\ref{copleibfunvar}).

\subsubsection*{Acknowledgments}
We thank  Francesco Bonechi, Giuseppe Marmo and Harold Steinacker
for useful discussions and correspondence. 
We thank Julius Wess for his encouragement and advice concerning this work, 
that has been completed just when he passed away.
We acknowledge hospitality and partial support from the Erwin Schr\"odinger
Institute where this project started, and from II. Institut f\"ur Theoretische 
Physik, U. Hamburg under DFG programme SPP 1096, where it has been finished.
Partial support form the Quantum Geometry and Quantum Gravity
Programme of the European Science Foundation, from A.v.Humboldt foundation, and
from the European Community's Human 
Potential Program under contract MRTN-CT-2004-005104 and the
Italian MIUR under contract PRIN-2005023102 is also acknowledged.

\end{document}